# The Non-uniform Fast Fourier Transform in Computed Tomography (No. 2.4.4)

**Masters Report**

## Supervisor: Prof. Mike Davies

**Billy Junqi Tang s1408760**

UNIVERSITY OF EDINBURGH

# Abstract

This project is aimed at designing the fast forward projection algorithm and also the backprojection algorithm for cone beam CT imaging systems with circular X-ray source trajectory. The principle of the designs is based on utilizing the potential computational efficiency which the Fourier Slice Theorem and the Non-uniform Fast Fourier Transform (NUFFT) will bring forth.

In this Masters report, the detailed design of the NUFFT based forward projector including a novel 3D (derivative of) Radon space resampling method will be given. Meanwhile the complexity of the NUFFT based forward projector is analysed and compared with the non-Fourier based CT projector, and the advantage of the NUFFT based forward projection in terms of the computational efficiency is demonstrated in this report.

Base on the design of the forward algorithm, the NUFFT based 3D direct reconstruction algorithm will be derived. The experiments will be taken to test the performance of the forward algorithm and the backprojection algorithm to show the practicability and accuracy of these designs by comparing them jointly with the well-acknowledged cone beam CT operators: the CT linear interpolation forward projector and the FDK algorithm. This Master report will demonstrate a novel and efficient way of implementing the cone beam CT operator, a detailed summary of the project, and the future research prospects of the NUFFT based cone beam CT algorithms.

**Declaration of Originality**

I declare that this thesis is my original work except where stated. This thesis has never been submitted for any degree or examination in any other university.

……………………………………….

Junqi Tang

13/08/2015

# Contents

## Introduction

The predissertation report of this project [1] presented a very detailed literature review and explanation of the background theories for developing Fourier based cone beam CT (for circular trajectory) algorithms including the cone beam CT geometry, the Radon Transform, the Fourier Slice Theorem, the Tuy's sufficiency condition, the Grangeat relationship, and the Non-uniform Fast Fourier Transform algorithm.

We started the predissertation report by introducing and discussing the Radon transform and Fourier Slice Theorem (both 2D and 3D cases). The Radon transform provide us with the mathematical description and modelling of parallel CT projection, and the Fourier Slice Theorem provide us a link between the Radon transform of the image and its Fourier domain. If we can perform the reconstruction process (or forward projection process) of the image in the Fourier domain, the computational efficiency of the Fast Fourier Transform (FFT) can be utilized to develop fast forward and backward cone beam CT operators.

Next, the cone beam CT geometry and the Tuy's sufficiency condition [2] is discussed. Because in practice the circular cone beam CT geometry is playing the dominant role, this Masters project focuses only on this geometry. According to Tuy's sufficiency condition, the circular source trajectory will always have a missing data problem, which will cause the imaging system underdetermined.

Another important theoretical foundation is the link between the cone beam projection data set and the $1^{st}$ order derivative of Radon Transforms (or rather $1^{st}$ order derivative Radon space), and this is the Grangeat relationship [3]. This relationship enable us to process the cone beam projection data to the umbrella-like sampled (derivative of) Radon space.

In 1993, C. Axelsson [4] [5] gives an FFT based fast operation of Grangeat formula, which can tremendously reduce the computation demand of this operation. In this project, a similar but advanced idea is used to execute the Grangeat cone beam data processing procedure.

In the section 2.7 of the pre-dissertation report, the theoretical foundation of Non-uniform Fast Fourier Transform has been laid in detail. In this section the structure of the NUFFT and its transpose algorithm is introduced. Then the terms of approximation errors in discussed and explained including aliasing error and truncation error. After these, I introduced a widely acknowledged NUFFT algorithm given by J. Fessler [6], which is optimal in the min-max sense. This means that the maximum possible error for every single point at the output is minimized, which is a very essential property in Fourier domain operation because if one frequency point have large error, it will give a big effect in the image domain.

The insights of applying NUFFT in the CT imaging algorithm is shown at the end of section 2.7 of the pre-dissertation report [1]. In this section a mathematical description of the general CT imaging system (forward direction), and comparatively, a mathematical description of the NUFFT based cone beam CT forward operator is given. What follows is a discussion about the CT imaging system's inversion problem, and we can see in the direct reconstruction algorithm a ramp filter is needed to correctly reconstruct the image without any discrepancy. For a cone beam backward imaging operator, a direct ramp filter is not found.

There are now two approaches of dealing with the absence of the cone beam system's ramp filter. The first one is using the iterative image reconstruction algorithm, as shown in the section 2.7 of the pre-dissertation report, if we choose to solve the imaging system iteratively, the ramp filter for cone beam geometry is avoided, what we need is only the forward operator and its transpose algorithm. The second idea is to utilize the fact that the ramp filter for the 3D parallel projection is known as '$R^2$' in the Fourier domain, if one can perform a rebinning from the cone beam projection data to the parallel projection (or the radially sampled 1st order derivative of Radon space), then the already known 3D ramp filter can be utilized and then we can solve the cone beam system directly, which means the reconstructed image will be theoretically data-consistent.

With these established foundation, the main assignment of the project is to design the NUFFT based forward projection algorithm for cone beam geometry, and also to design the NUFFT based direct image reconstruction algorithm for circular cone beam CT system.

The first step of this stage of the research is to discuss and determine the angular sampling rate we need to used when apply the Fourier Slice Theorem in the algorithms which are going to be designed in this project, and this will be the content of the chapter 1.

The next part of the research (in chapter 2) is the designing of the forward projection algorithm, which should generate cone beam projection data from a given 3D image. In this step the forward operator which was very briefly described in the third chapter of the pre-dissertation report will be established. The most essential part of the chapter is the (derivative of) Radon space resampling method. The classic resampling method will be introduced. But in this project a novel 3D space resampling method is derived, which is initially designed for the forward direction (resampling from radial sampling pattern to umbrella sampling pattern). The resampling method proposed in this report has two variations, the first one is NUFFT based while the second one only utilize the interpolation matrix of the NUFFT algorithm. After finishing the design of the forward operator, in section 2.4 we will give a complexity analysis of the NUFFT forward projector and the CT projector based on the linear interpolation principle, and discuss the computational advantage the NUFFT forward projector will bring forth in cone beam CT.

Using the same principle we can derive the direct reconstruction algorithm from the forward algorithm as well, and this design is given in section 2.3. Note that the essence of the designing this direct reconstruction algorithm is to resample the (derivative of) Radon space into radial sampling pattern, then the correct ramp filter for this particular sampling pattern can be used to reconstruct the 3D image.

But in this project the resampling method is derived rigorously only in the forward direction, if we use it reversely, this step will theoretically introduce an extract error. But from the experimental results we can see that although this method is not rigorous in the resampling step, it is able to reconstruct the image in a reasonable accuracy and has potential to be treated as an alternative option for the widely applied FDK algorithm [7] in the future after the further research and improvement.

The third chapter of this report will focus on the experiments and tests of the NUFFT based forward algorithm with both two resampling methods, and direct reconstruction algorithm, while the first experiment is designed to empirically demonstrate the reliability of the sampling rates which is going to be used when applying the Fourier Slice Theorem. Next, the NUFFT forward projector with these sampling settings will be implemented, and then the projection data generated by the NUFFT projector will be compared with the projection data generated by the CT projector given by Kyungsang's Matlab toolbox, and see the difference between them by observing the difference images. The third is to test the backprojection algorithm designed in this project by reconstruct the 3D Shepp-Logan phantom image from 1) the projection data generated by the NUFFT forward projector, and 2) the projection data generated by the CT projector, and a comparison with the FDK algorithm's results is also given. This experiment will not only demonstrate the performance of the fast backprojection algorithm but also demonstrate the accuracy of the NUFFT forward projector because the well-acknowledged FDK algorithm is used to reconstruct the 3D phantom from the projection data given by both NUFFT projector and CT projector.

The fourth chapter will give a summary of the progress of the research and the further prospects.

# 1. The theoretical background for Radon space (Sinogram) sampling

This chapter will investigate the application of the Fourier Slice Theorem, ramp filtering and NUFFT algorithm.

In Fourier based CT imaging, the Non-uniform FFT algorithm and the Fourier Slice Theorem are the rock, as already discussed in my pre-dissertation report. The Fourier Slice Theorem gives the parallel beam projection data direct access into the Fourier domain, and results in Non-uniform frequency samples; while NUFFT provides a fast and well-approximated computation from the Fourier samples to the image we want. This chapter will go further from the basic of the theory towards the details in actually implementing it, discover and solve the problems we may face in practice.

The first question we need to answer is that, for reconstructing a certain size (say, N*N) of image, how many frequency sampling lines do we need in the Fourier domain. Due to the sampling geometry demanded by the Fourier Slice Theorem, the sampling in the Fourier domain is inefficient (intense sampling in low-frequency and lesser sampling in high frequency).

In order to get a rigorous discussion, let us establish the basic assumptions of the 2D continuous signal we are dealing with:

1) It is reasonable to assume that the 2D object has a round edge (please imagine that in the fan beam projection case, the full-projected area----which can be covered by all the projections must be shaped as a round). The assumption means that the object is space limited.
2) The second easily accepted assumption is that the 2D signal is bandlimited in 2D Fourier domain in a certain radial distance $\boldsymbol{R}$ from the DC component. In other words we assume that the high frequency components which are located outside the range $\boldsymbol{R}$ can be approximately regarded as zero. This assumption is an inherent default assumption in CT imaging research.
3) The third assumption is that the sampling rate of the target image satisfies the Nyquist Sampling Theorem, which means that by this sampling rate we can capture all the information of the 2D continuous signal.

The mathematical proof of the minimum angular sampling criterion can be found in [8]. At this chapter we will illustrate the way to determine the minimum number of sampling radial lines in the 2D Radon space for a given 2D continuous signal which can be sampled by an N by N sized image without losing information, and in chapter 3 there will be a numerical experiment designed to empirically demonstrate this.

The figure 1.1 illustrates the space-limit assumption and the band-limit assumption of the 2D continuous signal, and the spectrum of the 2D signal's Sinogram (2D Radon space).

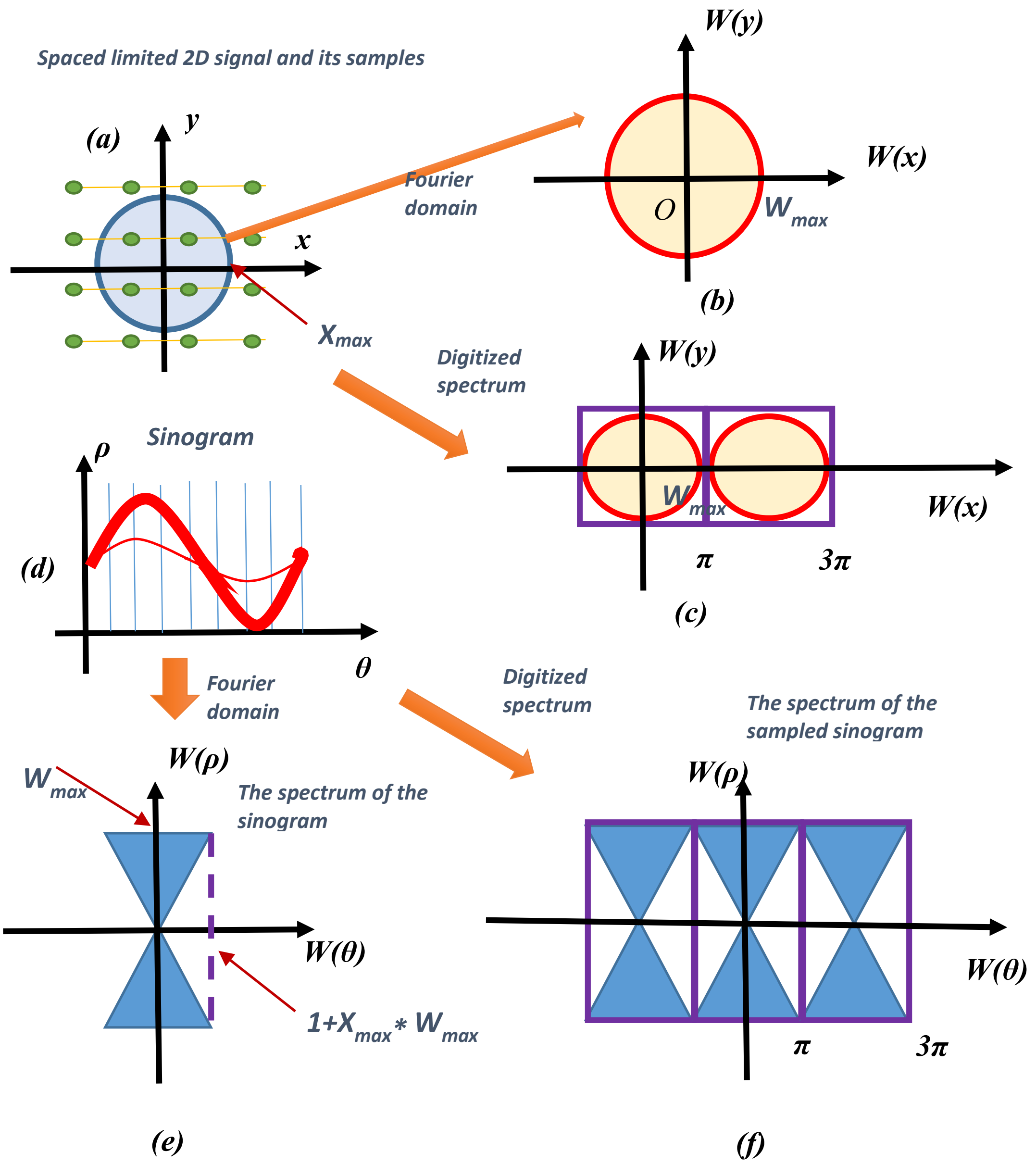

***Figure 1.1 The illustration of how the 2D Sinogram sampling is related to the original image's bandwidth and how to correctly sample the 2D Sinogram without aliasing***

As shown in figure 1.1(a) and 1.1(b), the input 2D continuous signal is space-limited in a circle with the radiance $X_{max}$, and is also bandlimited in a circle with the radiance $W_{max}$, and the values outside the circle is assumed to be negligible.

According to Nyquist Sampling Theorem, if we sample the signal by a rate:

$$\Delta x = \Delta y = \frac{\pi}{\mathrm{Wmax}} \tag{1.1}$$

$$\mathrm{Nx} = \mathrm{Ny} = \frac{2Xmax}{\Delta x} = \frac{2\mathrm{Xmax}*\mathrm{Wmax}}{\pi} \tag{1.2}$$

(Nx and Ny are the numbers of samples in x and y direction respectively.)

The information of the 2D signal is perfectly preserved, as shown in figure 1.2(c), there is no aliasing occurred in frequency domain.

Then let us now investigate the Sinogram and its 2D Fourier Transform (treat the Sinogram itself as a 2D signal in ρ and θ), interestingly every Sinogram has a bowtie-shaped spectrum, and the highest frequency for ρ is $W_{max}$, and the highest frequency for θ is $1+X_{max}*W_{max}$, according to [8]. Obviously:

$$\Delta\rho = \frac{2\pi}{2\text{Wmax}} = \frac{\pi}{\text{Wmax}} \tag{1.3}$$

$$N\rho = \frac{2Xmax}{\Delta\rho} = 2 * \frac{\text{Xmax}*\text{Wmax}}{\pi} = Nx \tag{1.4}$$

$$\Delta\theta = \frac{2\pi}{2(\text{Xmax}*\text{Wmax}+1)} = \frac{\pi}{\text{Xmax}*\text{Wmax}+1} \tag{1.5}$$

$$N\theta = \frac{2\pi}{\Delta\theta} \cong 2\text{Xmax} * \text{Wmax} = \boldsymbol{\pi} Nx \tag{1.6}$$

(Nρ and Nθ are the numbers of samples in ρ and θ direction respectively.)

**So from the results above we can conclude that the sampling rate for a 2D 128 by 128 image's Sinogram against (ρ, θ) is Nρ=128 and Nθ greater than 128π; the sampling rate for a 3D 128-cube-sized 3D image's Sinogram against (ρ, θ, φ) should be Nρ=128 , Nθ and Nφ both greater than 128π.**

But in the experiments of this project this sampling rate setting is not used. Because of the time limit of the project, a sampling rate which is much easier to implement is used instead: when sampling the 128-cube-sized 3D image's Sinogram (this is the first step of the NUFFT forward projector), both Nρ, Nθ and Nφ are chosen to be 256, which means that we oversample ρ by a factor of 2, and reduce the sampling rate Nθ and Nφ from 128π to 128*2=256, which is a reasonable compromise because this sampling rate will give promise the enough sampling number at ρ and θ direction, but will lose some information at φ direction's high frequency part. In the future research, an accurate experiment should be taken to implement and test the NUFFT projector strictly follows the sampling rates given by the theoretical derivation.

From the 2D case's result given by (1.6) we can see that theoretically the minimum number of equal spaced Sinogram sampling against θ is $\boldsymbol{\pi N_x}$, which is quiet a large number, especially for a 3D imaging problem. But from the spectrum of the uniformly sampled Sinogram illustrated in the figure 1.1(f), we can see that the Fourier space is not filled up efficiently, a half of the space is wasted, which implicitly tells us that a half of the samples we take is not necessary.

If we use hexagonal sampling pattern instead of the uniform sampling pattern to sample the Sinogram, the angular sampling can be reduced by a factor of 2 according to [8]. But this sampling pattern is difficult and complex to implement in the algorithms which are designed this project and need further research. Nevertheless we can utilize this property to do a trade-off between the sampling number in ρ-direction and the angular sampling number, as suggested by figure 1.2.

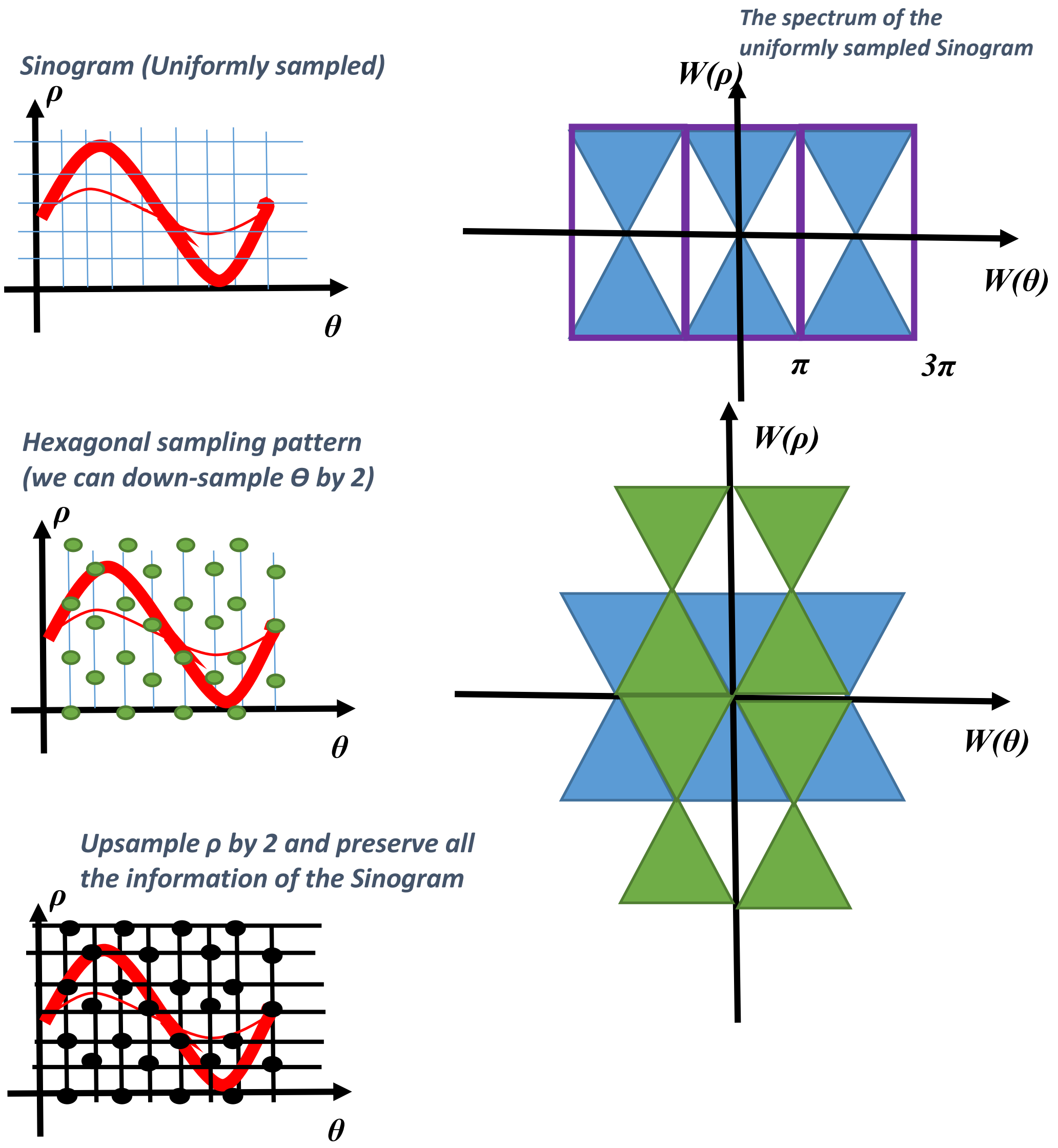

***Figure 1.2 The illustration of the efficient way of hexagonal sampling the 2D Sinogram, and by utilize this we can down-sample the angle Θ by up-sample ρ***

The figure 1.2 demonstrates that we can use hexagonal sampling pattern to efficiently sample the Sinogram and reduce the minimum sampling number of Θ by 2 without losing any information. And if at the same time we would like to preserve the simplicity of the uniform sampling, we can simply up-sample the ρ by 2 and go back to operating the uniform sampling pattern.

These results can justify the sampling rate settings on $\rho$ and $\theta$ of the designed algorithms in this project when applying the 2D Fourier Slice Theorem. But in 3D case the sampling rate on $\varphi$ is not yet adequate in the experiments taken in this project.

In the third chapter, there will be an experiment on the actual angular sampling rate we need for a 2D slice of Shepp-Logan phantom, a real world medical image, and a random noise image. And it shows that for a 128*128 sized phantom, the 256 equal spaced radial lines sampling in Fourier space, (or rather, in Radon space) is more than enough to preserve the information of the image when the sampling number along each radial line is chosen to be 256 (2 time oversampling). This numerical experimental result is consistent to the results of theoretical discussions above.

## 2. The designs of the NUFFT based cone beam CT algorithms

In this section, a design of the NUFFT based forward cone beam CT projection operator with complexity of $O(N^3 \log N)$ will be given. The main modules of the fast forward projection algorithm can be described as:

1) Fast calculation of 3D derivative Radon space (Radially sampled) by applying 3D NUFFT and 3D Fourier Slice Theorem.
2) Resample the 3D derivative Radon space, from radial line sampling to umbrella-shaped sampling (correspond to cone beam geometry).
3) Fast backprojection from each set of the umbrella samples belongs to each projection to the projection data by using 2D NUFFT's transpose algorithm.

The key step of this algorithm is the design of the Radon space resampling procedure in the second step, and this is one of the innovative points of this paper and will be derived in detail in the coming section.

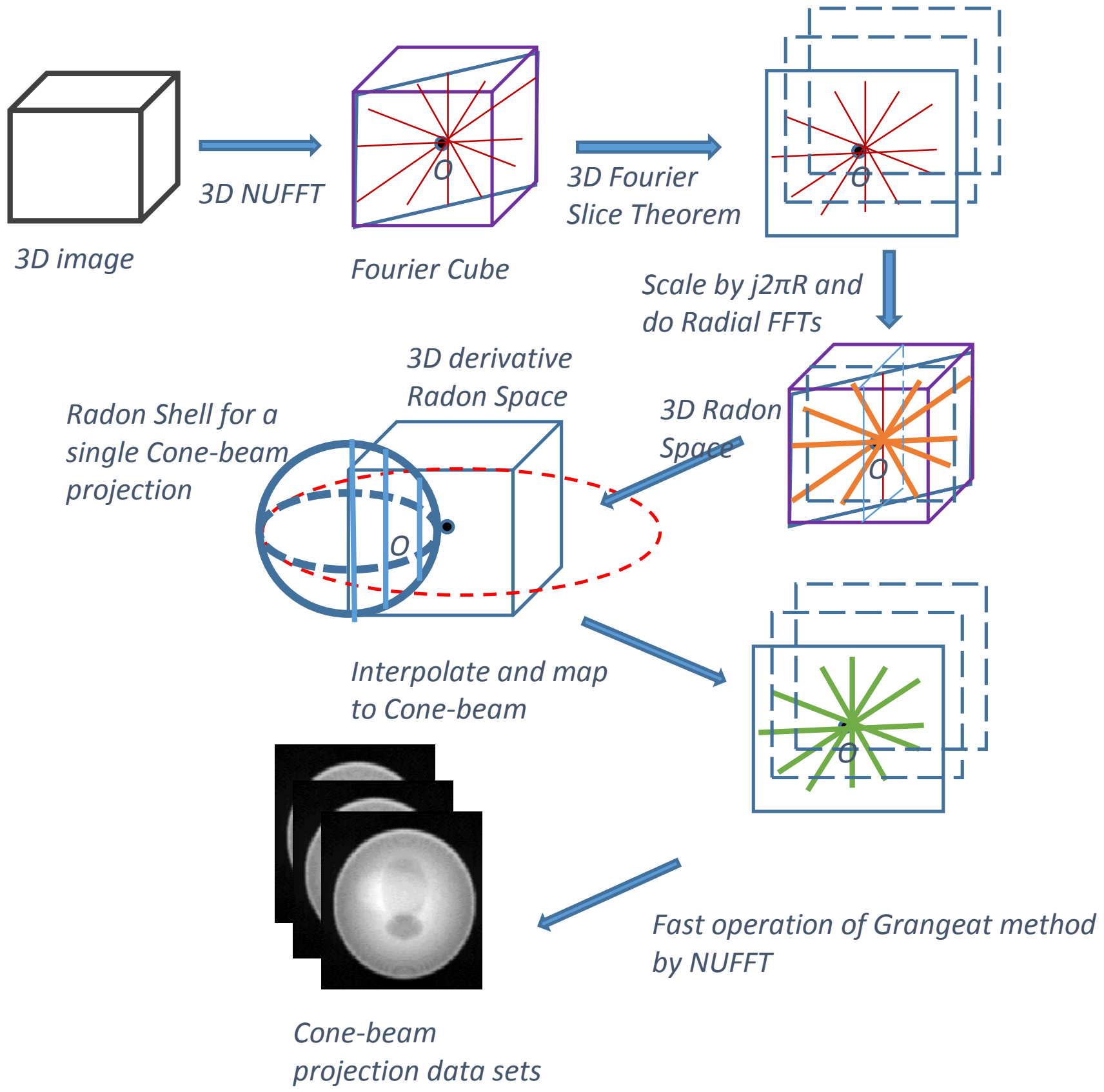

*Figure 2.1 The structure of the NUFFT based forward projection algorithm (modified from the predissertation report [1])*

## 2.1 The methods of Radon space resampling

### 2.1.1 Problem Statement

Let's look at the geometrical illustration of the problem we are facing:

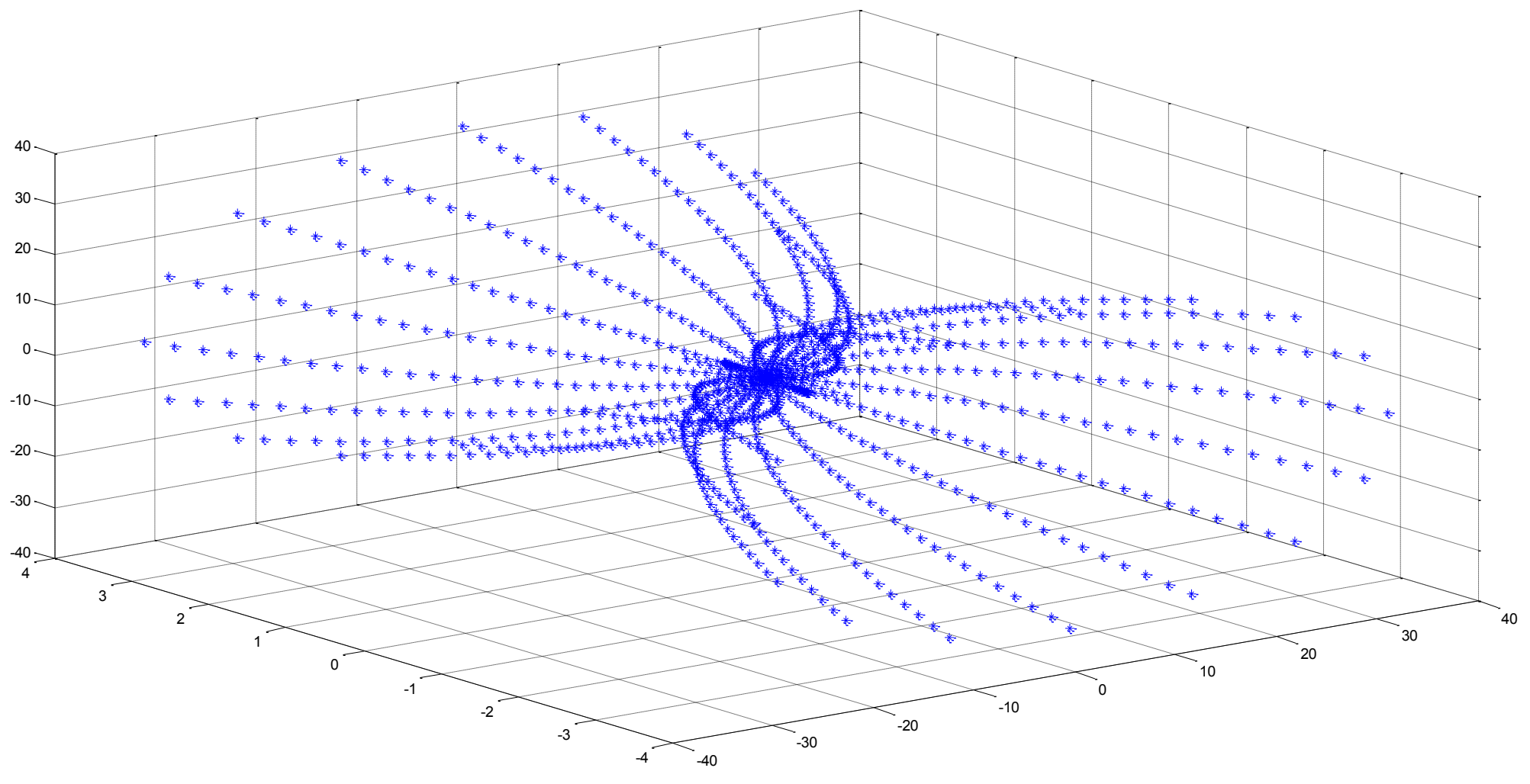

*Figure 2.2 The illustration of the umbrella-shaped Radon space sampling pattern according to cone beam geometry, for simplification, only the sample sets corresponds to two cone beam CT projections are shown here*

Figure 2.2 gives us an illustration of Radon space sampling related to cone beam projection geometry. But in order to get a fast computation of (the derivative of ) the Radon values from a 3D image, the Radon space should strictly be radially sampled, if not so, the Fourier Slice Theorem cannot be applied.

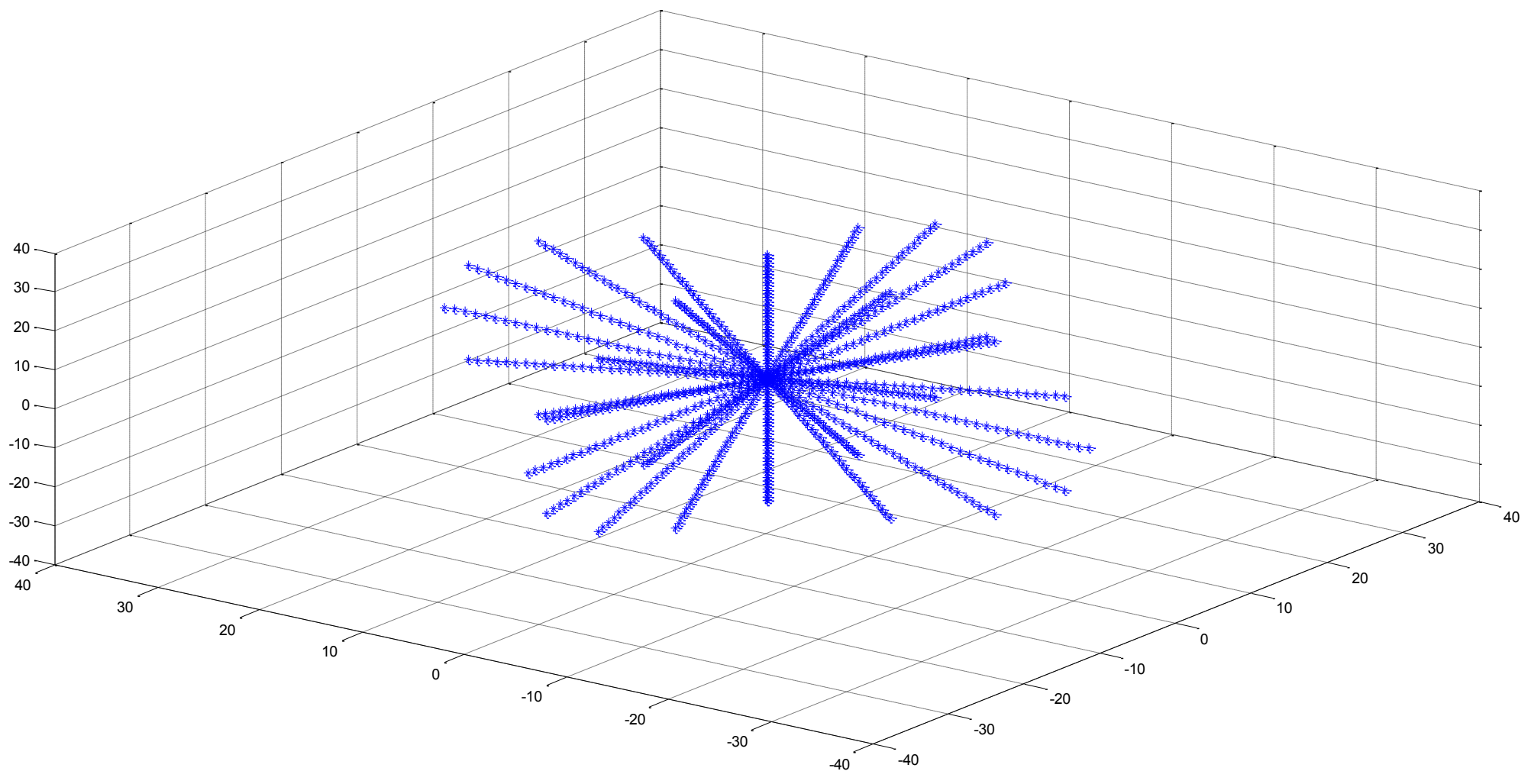

*Figure 2.3 The illustration of radial sampling pattern in the (derivative of) Radon space*

The purpose of this rebinning step in the NUFFT forward projector is to convert the sampling pattern of the (derivative of) Radon space from radial sampling to umbrella sampling. A traditional and intuitive way of performing the derivative Radon space resampling is the linear interpolation from nearest points. In C. Axelsson's 3D reconstruction paper [4], this method is chosen to be implemented.

The Cartesian positions of each points in radial sampling pattern can be easily calculated and described by knowing its azimuth and elevation angles. Assume the azimuth angle is φ, identifying the X-ray projection angle, the elevation angle is θ, and the distance towards the original point is ρ. The position of a point can be written as: [4]

$$x = \rho \cdot \cos\varphi \cdot \sin\theta$$

$$y = \rho \cdot \sin\varphi \cdot \sin\theta$$

$$z = \rho \cdot \cos\theta \qquad (2.1)$$

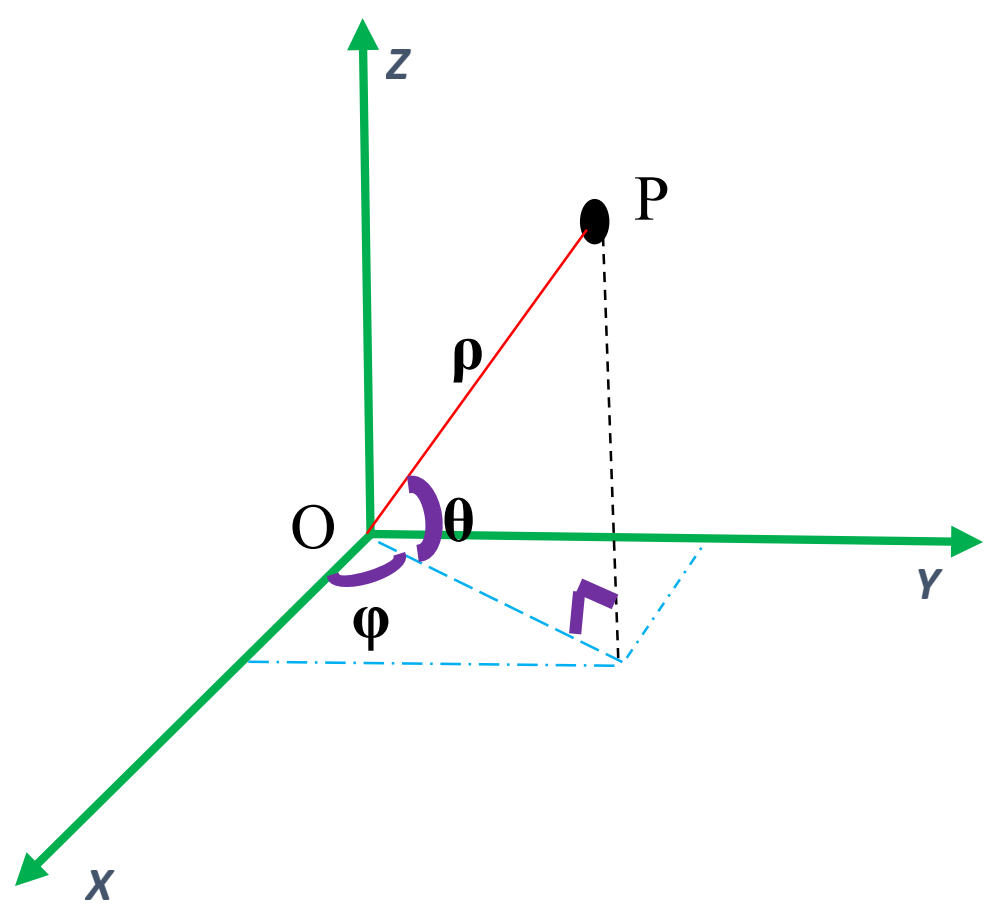

***Figure 2.4 The geometrical relationship between the Cartesian coordinate and polar coordinate, where φ is the azimuth angle, θ is the elevation angle and ρ is the geometrical distance between O and P***

The procedure of the linear interpolation in 3D space (given by C. Axelsson) can be described as:

Assume that we want to interpolate at a point with a random position in the radially sampled 3D (derivative of) Radon space.

(1) Find the nearest 8 points for this given position, as shown in figure 2.5.
(2) Calculate the distances between the targeted interpolation point towards each neighbouring points.
(3) Weight the (derivative) Radon values at each neighbouring points reversely by the distances calculated in step 2, and assign the summed-up value to the target interpolate point.

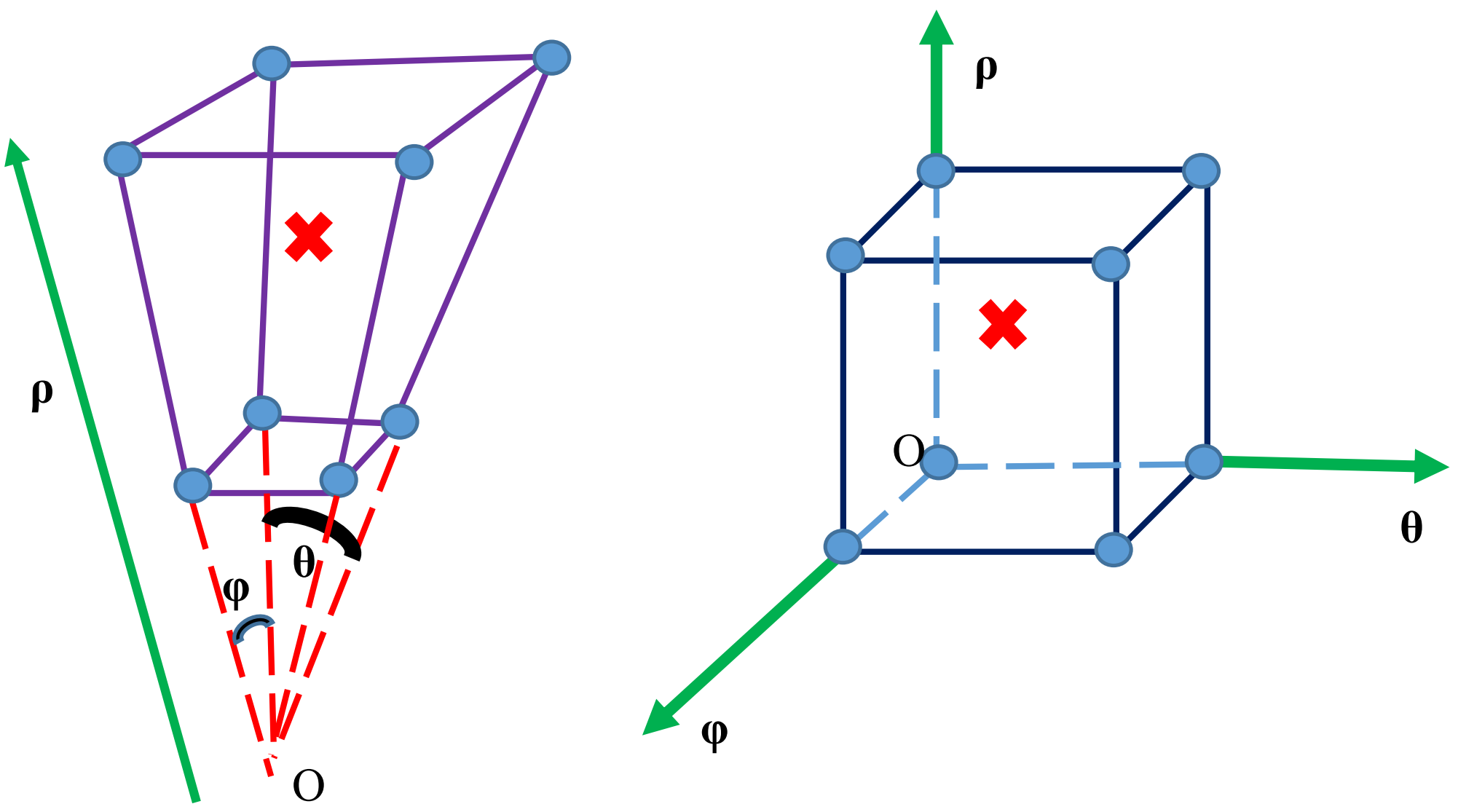

*Figure 2.5 The illustration of finding 8 nearest points (denoted by the blue points) for the targeted interpolate position (denoted by the red cross), where φ is the azimuth angle, θ is the elevation angle and ρ is the geometrical distance between O and P, the left figure represent the real geometry in the radially sampled 3D (derivative of) Radon space and the right figure present the positions of the points by polar coordinate, which is extremely useful in the coming session which is about the NUFFT based resampling method*

The linear interpolation is a reasonable way of performing this resampling, and according to C. Axelsson's paper, this interpolation is not computational consuming and produce reasonable results. But as one can see, this method is a practical way of solving the problem. From signal processing point of view, the filter's signal domain is triangularly shaped, as shown in figure 2.6.

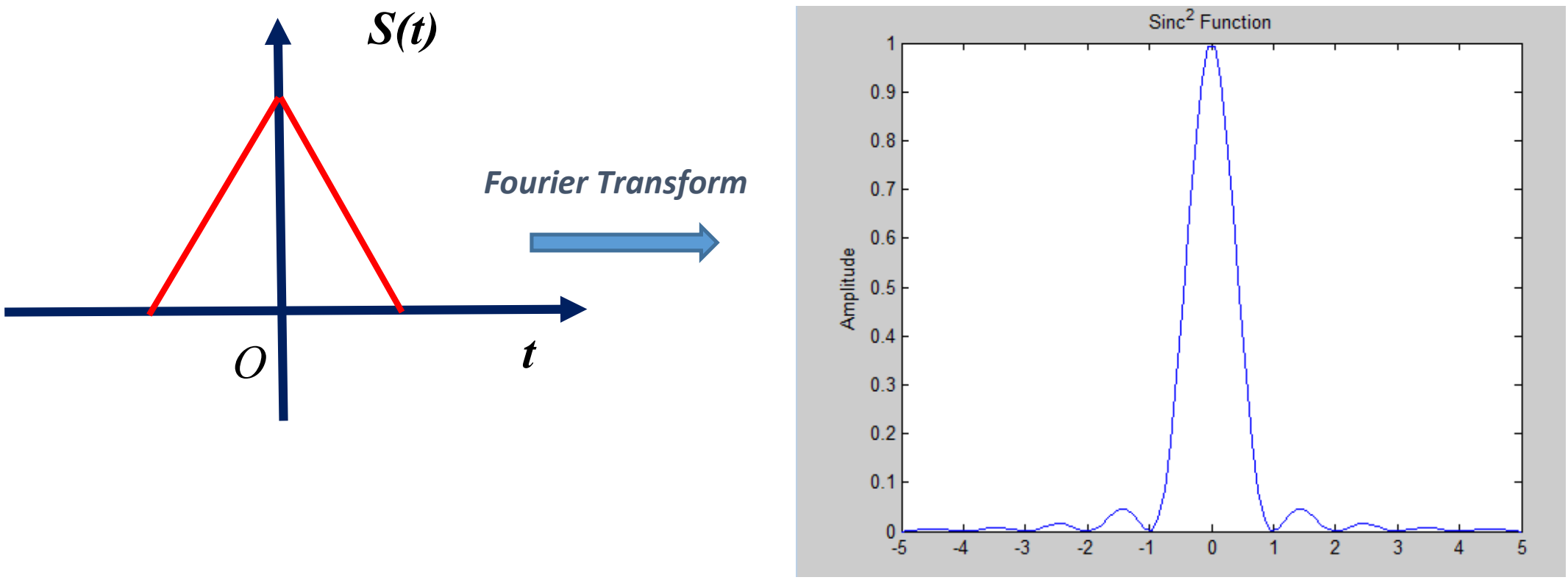

*Figure 2.6 The linear interpolation filter's signal domain and its frequency response*

This signal can be viewed as the convolution of two rectangular signal, so we can easily see that the frequency response should be the squared Sinc function. The frequency response of the triangular filter will have significant side-lobe, which means that the filtering performance may not be good, when we do subsampling, the aliasing will occur and introduce artifacts. And most importantly its frequency response is not flat at low frequency at all, so strictly speaking the linear interpolation is not a good choice and cannot suit the need for the 3D resampling step in this project.

In this paper a novel way of giving an interpolation which is implicitly and approximately doing ideal interpolation is going to be derived.

### 2.1.2 Interpolation and resampling theory

The optimal interpolation in terms of signal processing theory should be perceived in frequency domain. Let us assume a continuous signal which is unlimited in signal domain, and is bandlimited in frequency domain. Then sample the signal by a rate $F_s$ which is at least two times higher than the bandwidth $F_H$. By doing so, according to Nyquist Sampling Theorem, these samples preserve all the information of the original continuous signal and the perfect recovery is possible. If the signal can be perfectly recovered, then one can say that a perfect interpolation of a set of non-uniform samples from these uniform samples is also achievable.

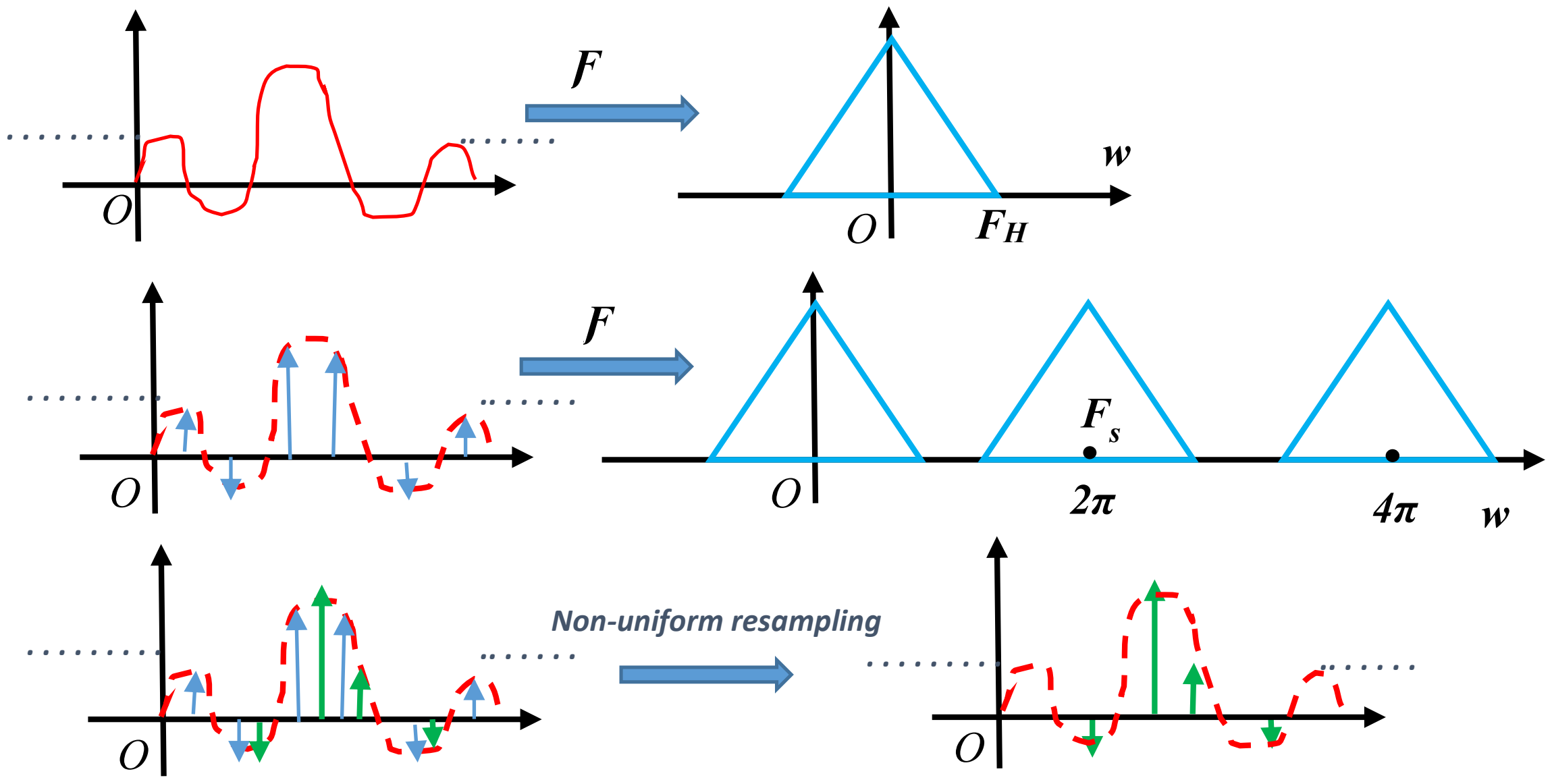

***Figure 2.7 The classic Nyquist digital sampling illustration, the example is: sampling a bandlimited continuous signal by a rate which is higher than twice the bandwidth of the signal***

Now we get the uniformly sampled points, the target is to optimally calculate the values in the non-uniform sampling positions by interpolation.

The optimal interpolation, or rather, optimal recovery of the original signal from the digital samples, as shown in figure 2.8, is to design an optimal low pass filter in the frequency domain, use this filter to optimally filter out the high frequency copies of the spectrum and preserve the spectrum which is belong to the original signal. The optimal filter's signal domain appearance is the sinc function, as the convolution kernel.

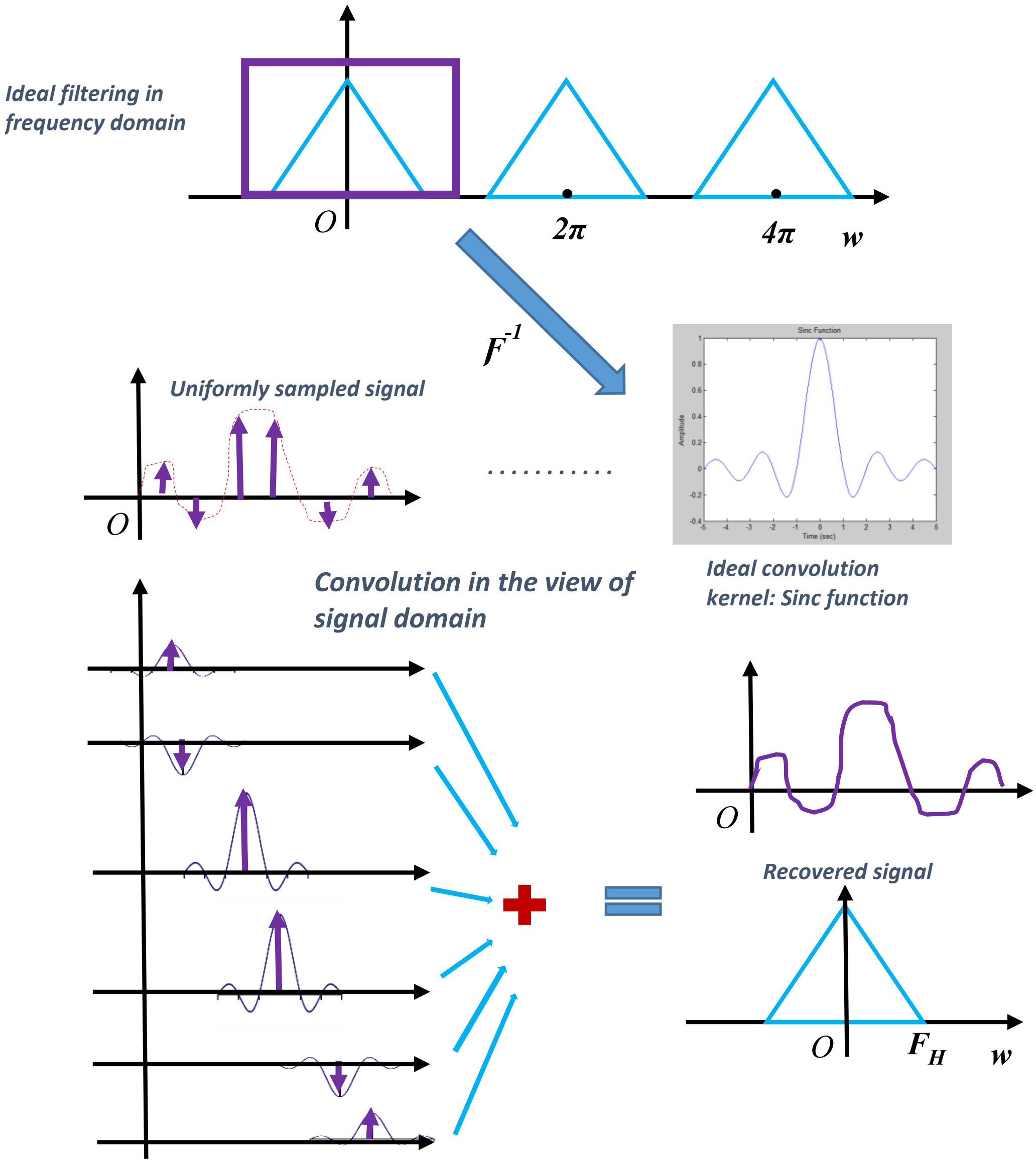

*Figure 2.8 An illustration of optimal recovery of a band limited signal by convolving with Sinc fuction*

Now we can also justify this interpolation mathematically from the Fourier domain of the signal, which gives light to a novel NUFFT based interpolation method. Assume that we have got a set of uniform samples y(k) of a band limited signal y(t), (k=1,2,3,4,5,…..), and the target is to calculate the values of the non-uniform samples y(m), where m is a non-uniform index (for example m=1.22, 2.56, 3.77…..), we can always calculate y(k)'s Inverse Discrete Fourier Transform (Note that N=K in this case since we do not need any oversampling):

$$x(n) = \sum_{k=0}^{K-1} y(k) e^{j\frac{2\pi}{N}nk} \tag{2.2}$$

And now obviously we can say that y(k) is x(n)'s spectrum. The non-uniform samples y(m) can be exactly represented by x(n) through using non-uniform Discrete Fourier Transform:

$$y(m) = \sum_{n=0}^{N-1} x(n) e^{-j\frac{2\pi}{N}mn} \tag{2.3}$$

Substitute (2.2) to (2.3):

$$y(m) = \sum_{n=0}^{N-1} \left\{ \sum_{k=0}^{K-1} y(k) e^{j\frac{2\pi}{N}nk} \right\} e^{-j\frac{2\pi}{N}mn} \tag{2.4}$$

Switch the order of summation:

$$y(m) = \sum_{k=0}^{K-1} y(k) \sum_{n=0}^{N-1} e^{j\frac{2\pi}{N}(k-m)n} \tag{2.5}$$

Solve the inner summation:

$$y(m) = \sum_{k=0}^{K-1} y(k) \frac{1-e^{j\frac{2\pi}{N}(k-m)N}}{1-e^{j\frac{2\pi}{N}(k-m)}} \tag{2.6}$$

$$y(m) = \sum_{k=0}^{K-1} y(k) \frac{e^{j\frac{\pi}{N}(k-m)N}}{e^{j\frac{\pi}{N}(k-m)}} \frac{\sin(\frac{\pi}{N}(k-m)N)}{\sin(\frac{\pi}{N}(k-m))} \tag{2.7}$$

Note that the term $\frac{\sin(\frac{\pi}{N}(k-m)N)}{\sin(\frac{\pi}{N}(k-m))}$ is the expression of a periodic sinc-kernel and the term $\frac{e^{j\frac{\pi}{N}(k-m)N}}{e^{j\frac{\pi}{N}(k-m)}}$ is simply a phasor. One can recognise that (2.7) expresses an "ideal" resampling from the uniform samples y(k) to non-uniform samples y(m), by using the sinc function as convolution kernel.

### 2.1.3 2D/3D optimal resampling from radial line sampling pattern

The obstacle we are confronting in extending 1D optimal filtering into 2D and 3D radial line sampling pattern is that (2.7) demands the sampling pattern is uniform so that the convolution kernel is shift-invariant.

But from [8] we can see that in 2D case if the number of radial line sampling are adequate enough (assuming the 2D continuous image signal is band-limited), the sampled Radon space still can uniquely determine the reconstructed image, which means, perfect information preservation. Because of this, the optimal filtering in the radially sampled Radon space is possible, the problem is how to approximately perform this optimal filtering by shift invariant convolution kernel in 2D.

Fortunately, there is uniformity hidden in this particular non-uniformly sampling pattern, which will enable us to execute the optimal filtering by simple shift invariant convolution. Note that although the radial sampling cannot be uniform in Cartesian grid, it is definitely uniform in polar grid. Because of this property, we need to apply the convolution in polar coordinate directions rather than traditional Cartesian coordinate directions.

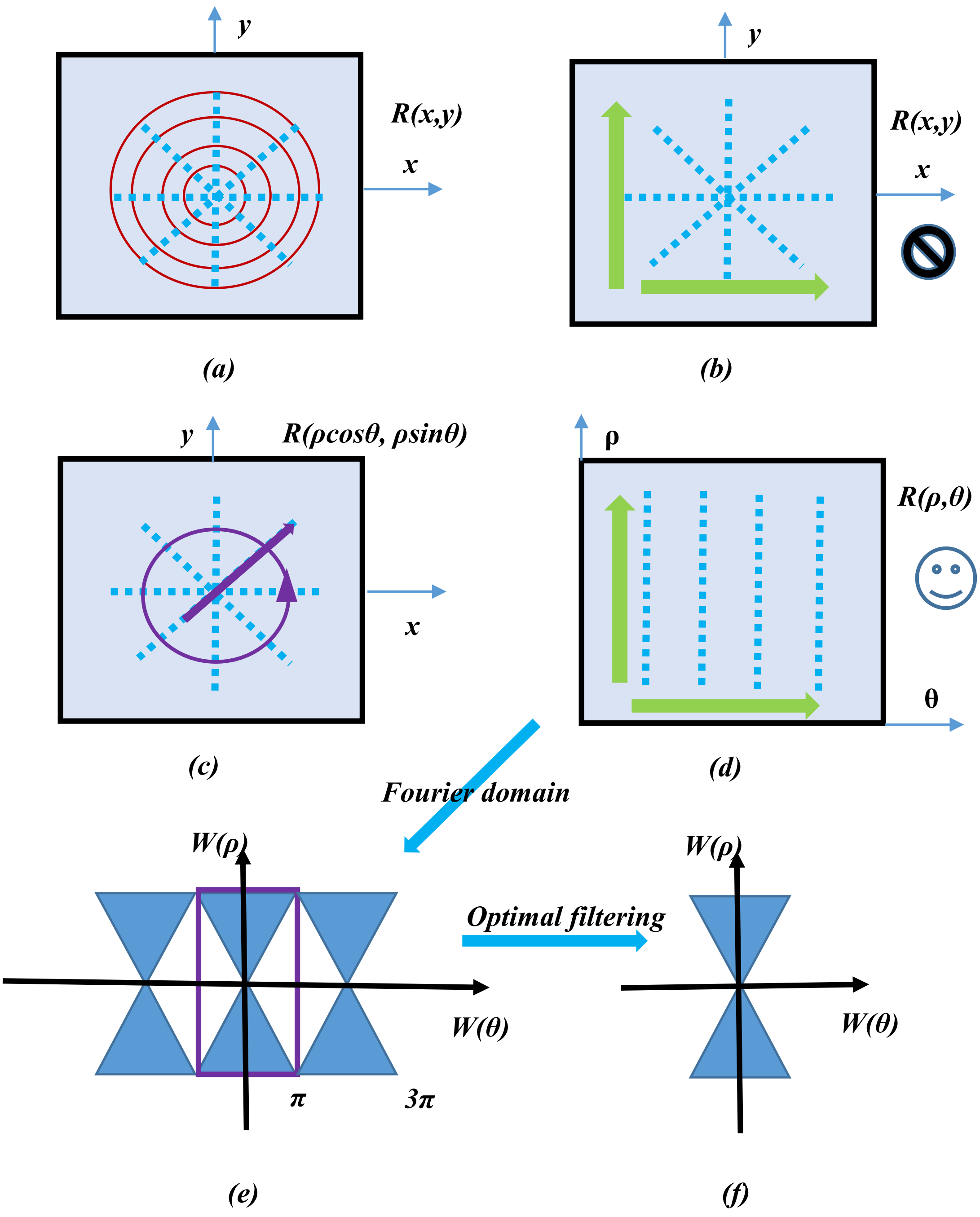

*Figure 2.9 (a) An example of radially sampled 2D Radon space. (b) Filter the radial samples along x and y direction, intuitively it will not work if we simply use shift invariant convolution kernels because the sampling density varies everywhere. (c) and (d) An alternative way of executing the filtering: angular filtering. (e) The spectrum of the 2D Sinogram. (f) Filtered spectrum of the Sinogram*

This idea of performing optimal 2D interpolation from radial sampling pattern can be easily extended to 3D case of Sinogram:

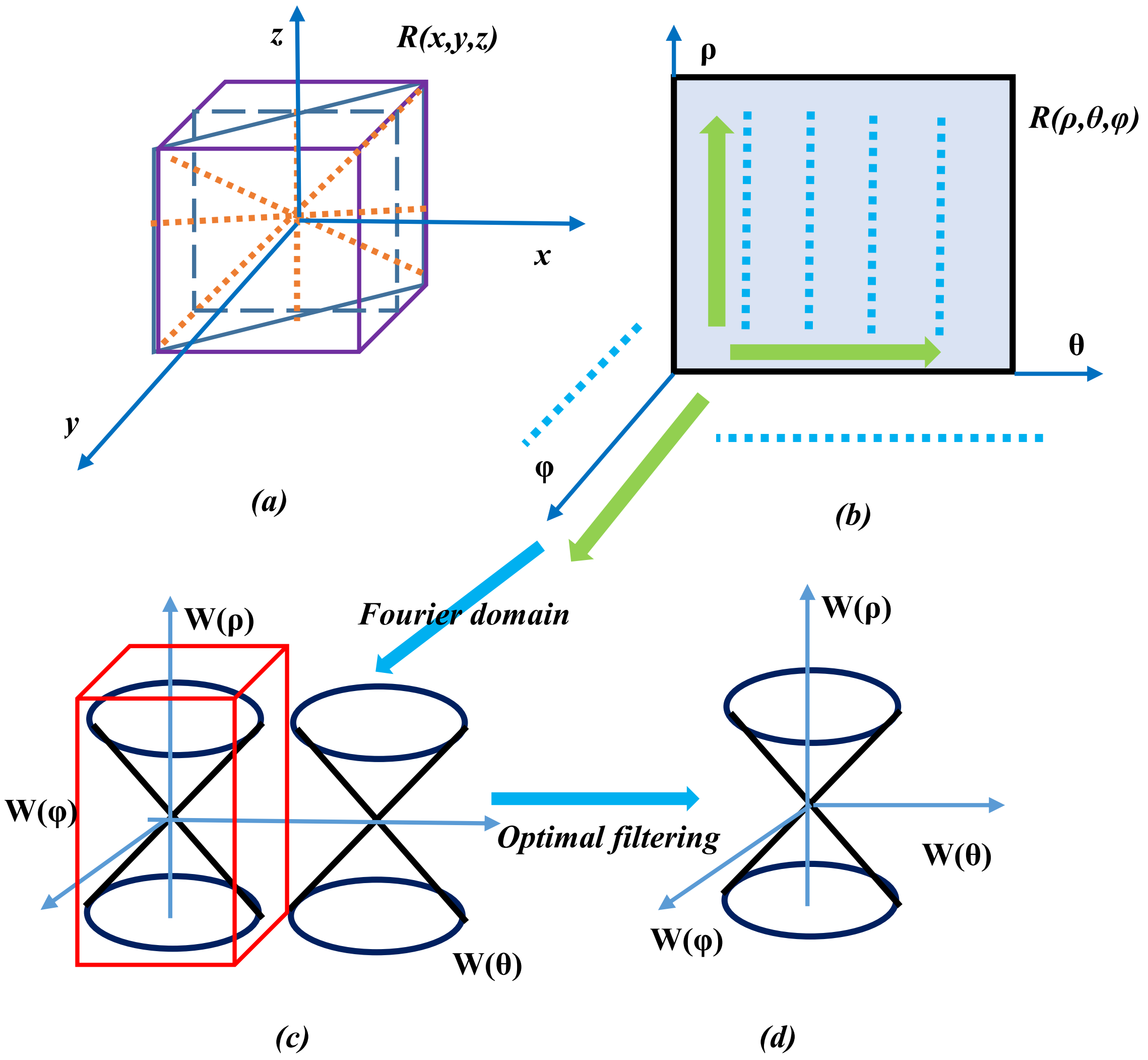

***Figure 2.10 (a) An example of radially sampled 3D Radon space. (b) Represent this space by polar coordinate (c) The spectrum of 3D Sinogram (d) The filtered spectrum of 3D Sinogram***

Now since we have already obtain a uniform representation in geometry for the radially sampled 3D (derivative of) Radon space, then we can go further into the derivation of NUFFT based interpolation method.

Assume we got a uniformly sampled and vectorised 3D signal $\boldsymbol{x}$ and the assignment is to interpolated this signal into non-uniform samples $\tilde{\boldsymbol{x}}$ **(Note that it is also a vectorised 3D signal)**, by an ideal interpolation operator $T$:

$$\tilde{\boldsymbol{x}} = \boldsymbol{T}\boldsymbol{x} \tag{2.8}$$

In (2.8), since we have made the assumption that T is an ideal interpolation operator, each row of T should contain a set of uniformly sampled value of a shifted and non-truncated sinc function, so T is a very non-sparse matrix. If x is vectorised from a 3D discrete signal sized N-cube, the operation of (2.8) should be $O(N^6)$, which is a very computationally inefficient way and cannot be applied directly in practice.

But we still desire to find a method which can well preserve the optimality of (2.8) and is able to be operated in a relatively fast speed. Let us do some expansion on (2.8):

$$\widetilde{\boldsymbol{x}} = \boldsymbol{T}\boldsymbol{F}\boldsymbol{F}^{-1}\boldsymbol{x} \tag{2.9}$$

Formula (2.9) is exactly equivalent to (2.8), where matrix F is the Fourier matrix, which can be operated swiftly by using FFT. But (2.9) does not perform any acceleration because the interpolation matrix T is still non-sparse. But if we insert an identity matrix between the Fourier matrix and inverse Fourier matrix, the opportunity to make T to be sparse will occur:

$$\widetilde{\boldsymbol{x}} = \boldsymbol{T}\boldsymbol{F}\boldsymbol{I}\boldsymbol{F}^{-1}\boldsymbol{x} \tag{2.10}$$

Now we can recognize that the term $\boldsymbol{TFI}$ is exactly NUFFT structure. This can be interpreted as a NUFFT with rectangular pre-windowing (or rather, pre-scaling), and T is the interpolator corresponds to the uniform pre-scaling operator, and this interpolator is exactly the sinc function, which is difficult to be truncated because of its relatively heavy tail.

But if we switch this NUFFT’s pre-scaling factors from uniform into a different form, for example, a raise cosine window, then the tail of the interpolator will be suppressed and then we can truncate it without facing the danger of losing too much optimality and make T a sparse matrix:

$$\widetilde{\boldsymbol{x}} \doteq \boldsymbol{T}_{\boldsymbol{Sparse}}\boldsymbol{F}\boldsymbol{P}\boldsymbol{F}^{-1}\boldsymbol{x} \tag{2.11}$$

Where P is a diagonal matrix and its diagonal contains the scaling factors. Formula (2.11) is the highly speed-up version of (2.8) with computation in the order of $O(N^3 \log N)$. Because the NUFFT with Kaiser-Bessel interpolation function promise a high quality approximation, this fast computation will be able to preserve the optimality of ideal filtering in a high extend and give a very reasonable trade-off between accuracy and computation.

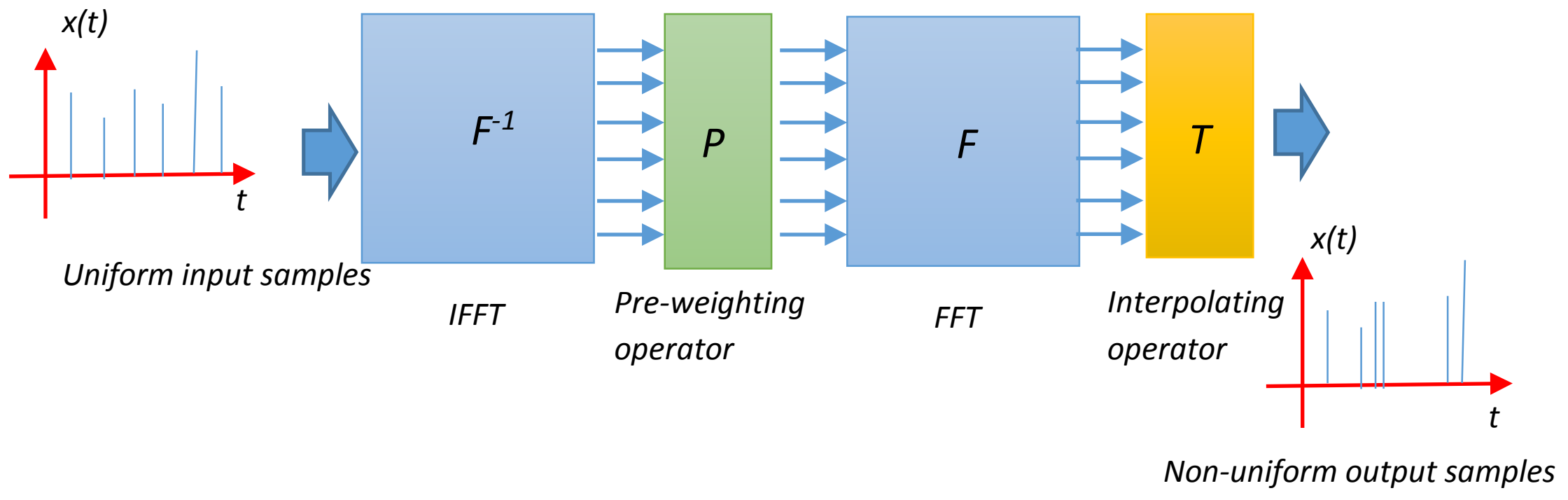

***Figure 2.11 the algorithm structure illustration of formula (2.11)***

If we would like to obtain the best accuracy in executing this resampling method, an oversampled NUFFT by a factor of 2 is preferred. This oversampling is rather essential if the Kaiser-Bessel interpolation kernel is used, because it is optimized when the oversampling factor is 2. But note that this oversampling is not necessary for all the available interpolation kernel and usually should be avoided in practice to save computation.

In the resampling method suggested by formula (2.11), the Fourier matrixes consume a part of the computation, although they can be executed by FFT. One may wish to speed up this resampling step further more by avoiding Fourier transforms, to get a deeper trade-off between accuracy and computation. Note that any Fourier matrix can be written as NUFFT-like structure:

$$\boldsymbol{F} = \boldsymbol{T_1 F I} \tag{2.12}$$

$$\boldsymbol{F^{-1}} = \boldsymbol{I F^{-1} T_2^T} \tag{2.13}$$

And we can substitute (2.13) into (2.10), and at the same time truncate the interpolation matrix T in to a sparse matrix by preserving the main-lobe and a few nearby side-lobes of the sinc function.

$$\widetilde{\boldsymbol{x}} \doteq \boldsymbol{T_{Sparse1} F I I F^{-1} T_{Sparse2}^T x} \tag{2.14}$$

Simplify formula (2.14), and a direct interpolation method is also derived, which contains two truncated sinc function operator:

$$\widetilde{\boldsymbol{x}} \doteq \boldsymbol{T_{Sparse1} T_{Sparse2}^T x} \tag{2.15}$$

Note that typically the second sparse matrix is an identity matrix. If the input data need reordering then this matrix should be a reordering matrix.

The resampling method given by formula (2.15) will give an alternative way compares to (2.11), and have potential to get a faster operation but at a cost of reducing accuracy.

## 2.2 The application of the Fourier Slice Theorem in the forward algorithm

### 2.2.1 The image-to-Radon transformation

In the former section we have got a detailed discussion of the resampling method for the 3D (derivative of) Radon space in the forward direction. Now let us look at the first step of the forward algorithm shown in figure 2.1.

The first step of the algorithm is the information transformation from the 3D image to the $1^{st}$ order derivative of the 3D Radon space. Note that as suggested in the predissertation report, this step by its nature is a very computation-consuming procedure because every single 3D Radon value is a result of a plane integration. In order to make this transform much faster, the 3D Fourier Slice Theorem is used as a link between the Fourier space and the (derivative of) Radon space, then the computational efficiency of the NUFFT may be used as a fast approach to calculate the Radon values.

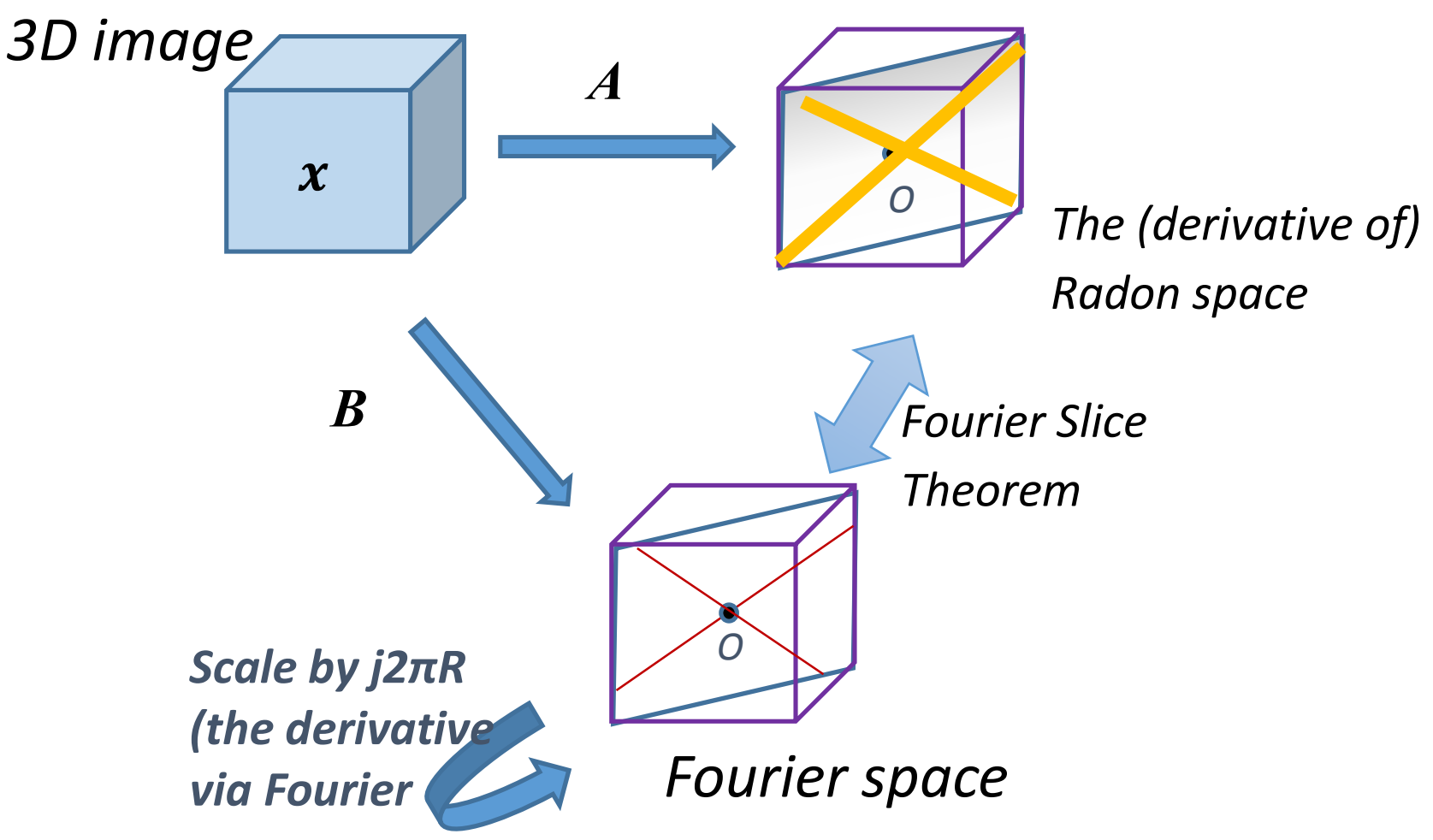

*Figure 2.12 The NUFFT and the Fourier Slice Theorem provide an efficient approach between the 3D image and the derivative of Radon space. In this figure, x is the 3D image, A is the expensive direct operator to transform the image to its derivative of Radon space, while B is the 3D NUFFT which will transfer the image to Fourier space where fast operation can take place.*

### 2.2.2 The fast implementation of Grangeat method in reverse

Since in the predissertation report I have already gave a detailed introduction of Grangeat method for processing the cone beam projection data to the first order derivative of Radon space, so in this Master report I will not repeat the same contents for detailed explanation again. Please note that the fast implementation of Grangeat method is initially given by C. Axelsson [4] in the backward direction for her backprojection algorithm by direct gridding in the Fourier domain, and in this project an improved version of this fast approach is implemented based on 2D NUFFT which can provide better accuracy in the Fourier space operation than the direct gridding.

The Grangeat method is initially derived for the backprojection algorithm instead of the forward projection, so in the designing of the forward operator we need to use the Grangeat relationship in reverse.

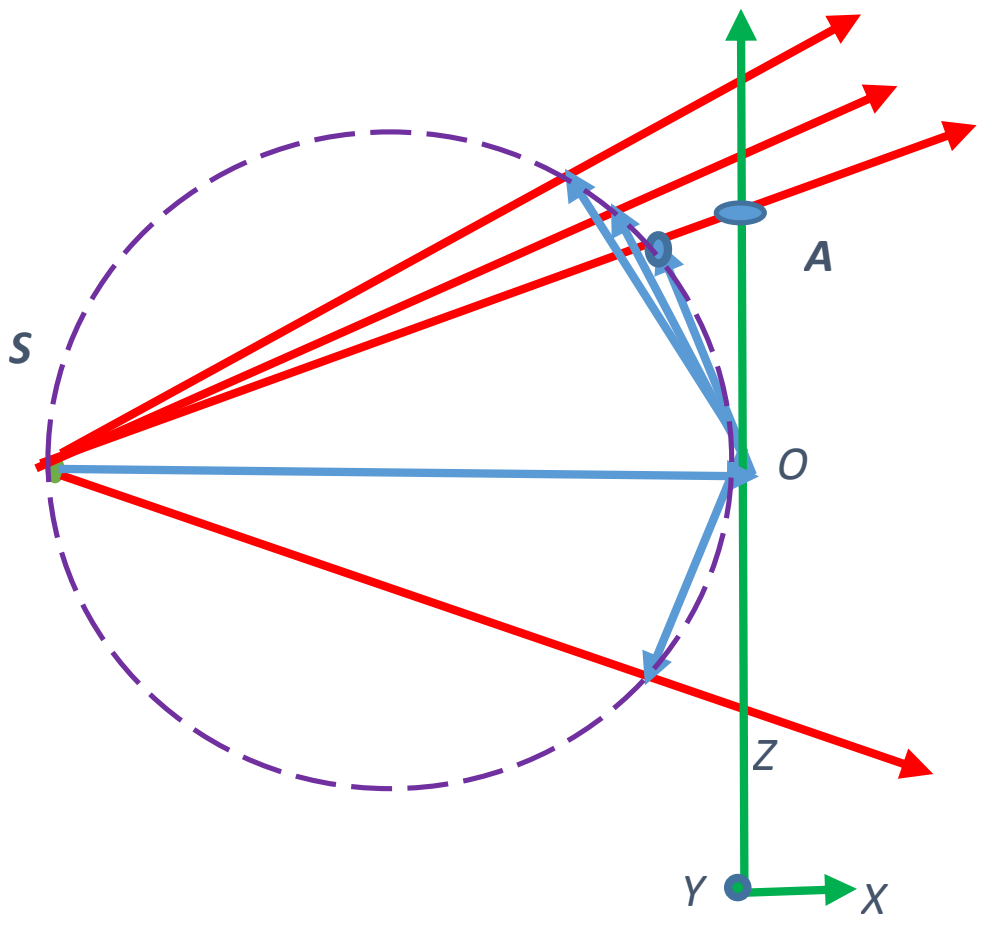

***Figure 2.13 The side-view of cone beam geometry, where S is the X-ray source, the detector plane is the Z-Y plane (this figure is from the predissertation report [1], it is placed here for the convenience of illustration)***

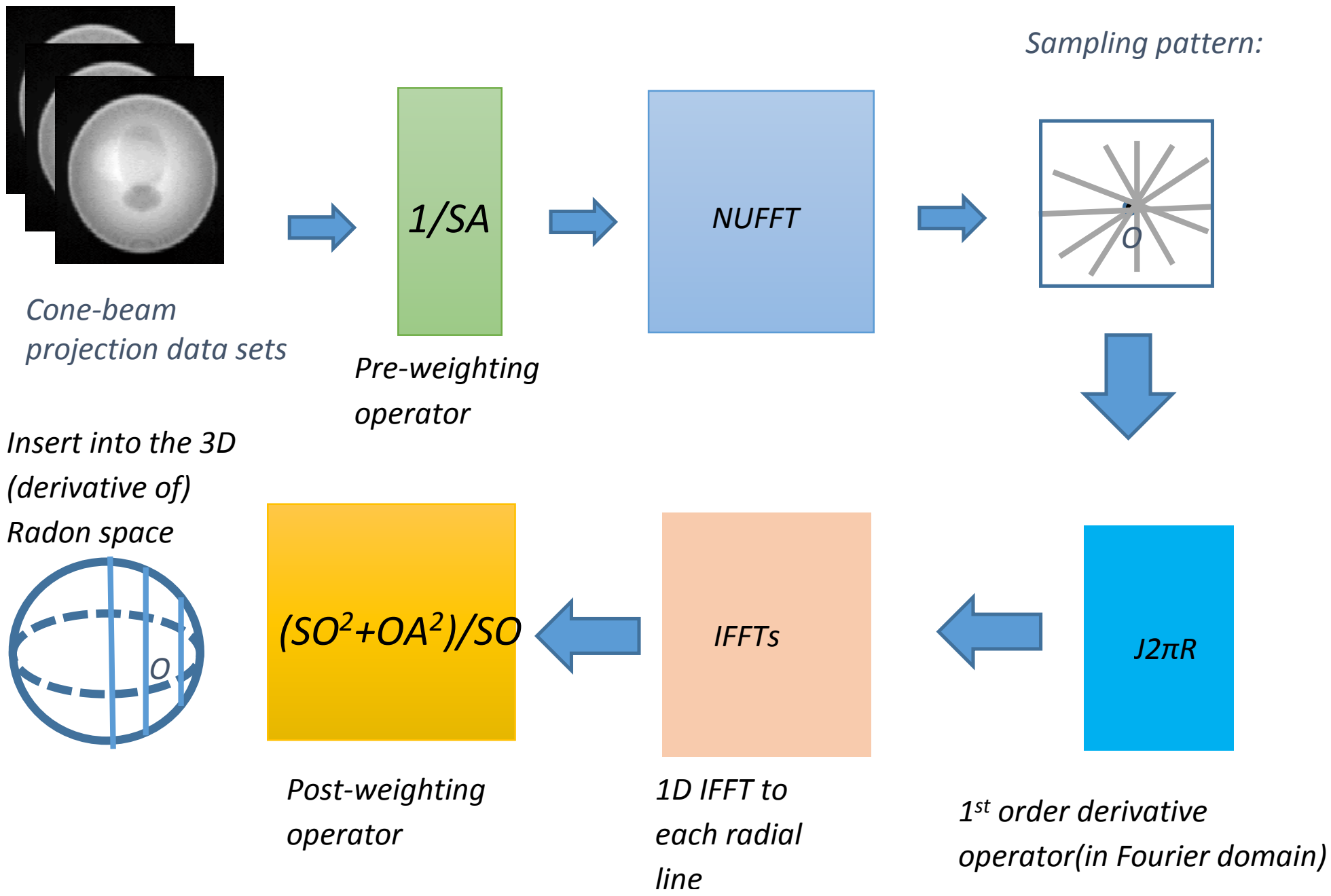

***Figure 2.14 The structure of the NUFFT based fast operation of the Grangeat method (it is in the backward direction)***

The figure 2.14 summarized the fast Grangeat algorithm's structure, and in the forward projection algorithm we execute the Grangeat method in reverse.

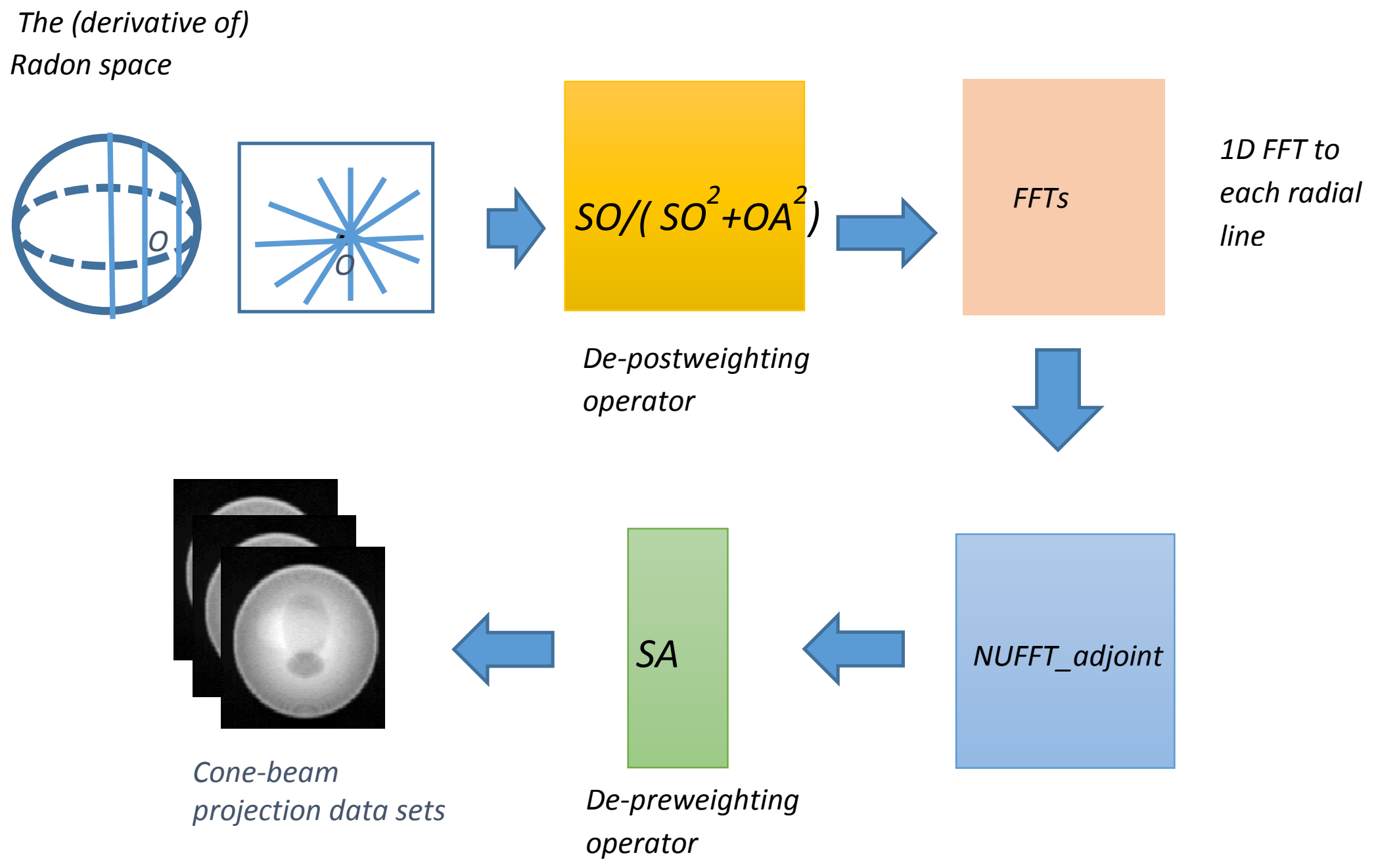

***Figure 2.15 The structure of the NUFFT based fast operation of the Grangeat method in reverse***

This reversed Grangeat method gives us a fast operation from the umbrella-sampled derivative of Radon space to a cone beam projection data set. As shown in figure 2.2, the umbrella-sampled Radon space is formed by many umbrella-like subsets of samples (each subset is uniquely related to one set of cone beam projection data). The algorithm suggested by figure 2.15 generate each cone beam data set separately (each time this algorithm is executed, one cone beam projection data set will be generated).

The first step of the algorithm is to take one umbrella-subset out of the 3D (derivative of) the Radon space, then just simply operate the Grangeat algorithm reversely, but note that the Radon values have already been taken the 1$^{st}$ order derivative, so at the NUFFT step there is no need for ramp filtering, we only need to use the NUFFT-transpose algorithm to do a 2D backprojection and get the correct results. Then repeat the algorithm to generate every set of the cone beam projection data.

From now on the detailed design of the NUFFT based forward projection algorithm is completed.

## 2.3 The NUFFT based direct reconstruction algorithm design

In the previous section, the NUFFT based forward algorithm has been designed, and in this section the idea of it will be utilized to develop a fast (filtered) backprojection algorithm for cone beam CT.

The fundamental scheme for designing this direct algorithm is to carefully inverse the forward algorithm, notice that most of the steps of the forward algorithm is completely invertible: the first step (the image-to-Radon transformation) can be inverted by a Fourier domain ramp-filtering operation and the NUFFT transpose; the third step's inverse is obviously the Grangeat method.

The only part of the forward algorithm which cannot be directly inversed is the 3D resampling algorithm. Because of time limit of the project, the transposed resampling algorithm is chosen to take the place of the inverse resampling.

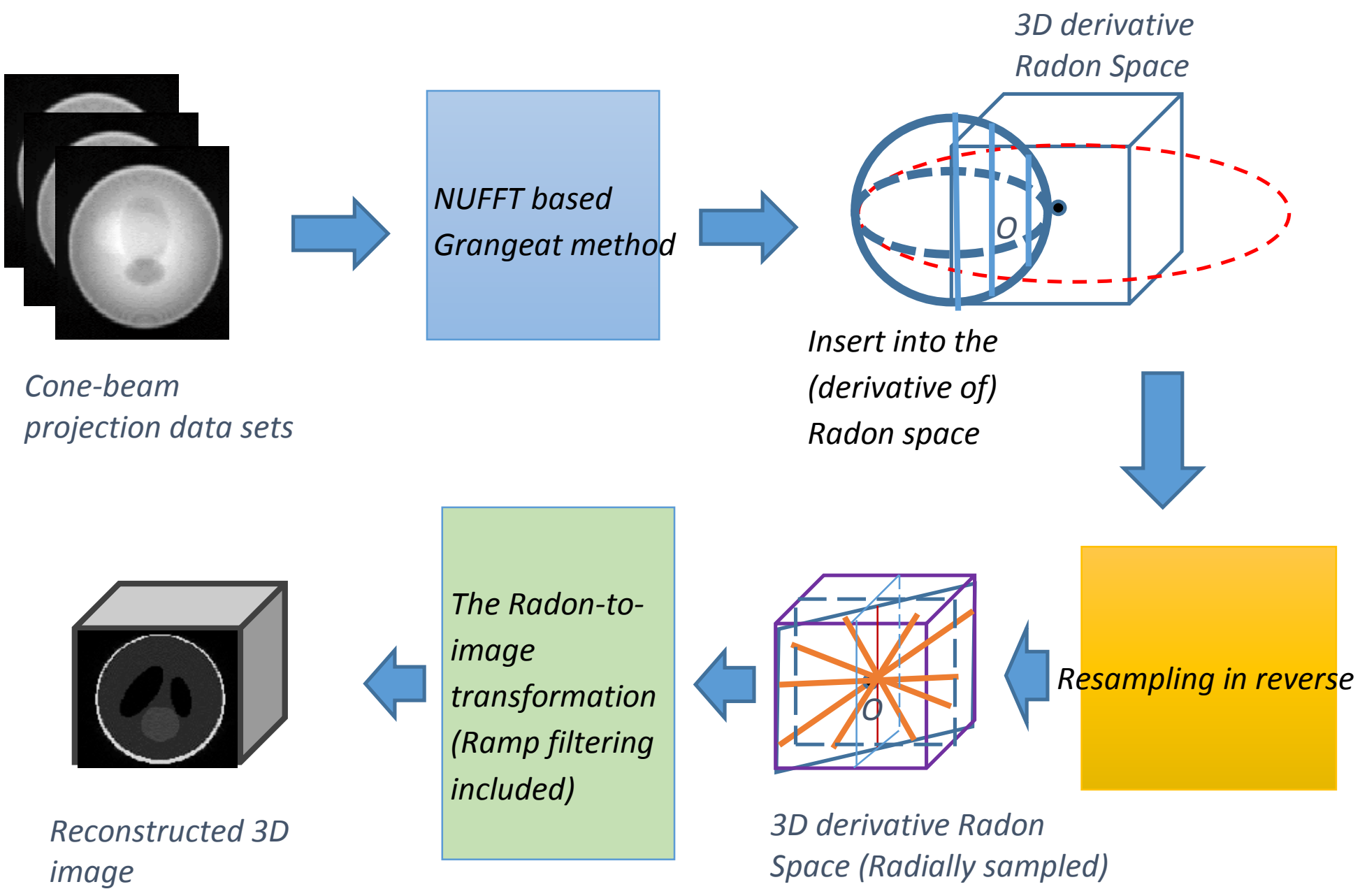

***Figure 2.16 The structure of the NUFFT (filtered) backprojection algorithm***

Since the discrepancy introduced by the resampling step, this direct reconstruction method is currently not a rigorous method. However from the experimental results of the algorithm, we will see that this discrepancy does not affect much about the reconstructed image and the algorithm will work at a reasonable accuracy.

## 2.4 The complexity analysis of the NUFFT forward projector and the CT projector and the comparison

This section's purpose is to give a calculation of the computational complexity of the forward projector and compare it to the complexity of the CT projection algorithm.

As described at the beginning of this chapter, the forward algorithm can be concluded as 3 steps: the image-to-Radon transformation, the 3D resampling, and the fast inversion of the Grangeat step.

Note that for the convenience of the analysis, this section only consider the number of multiplication, the summation operation is ignored since it consist a pretty trivial part of the computation when compared to multiplication operator.

For an N-sized 1D FFT, number of the 'butterfly-shaped' computation structure is ***0.25NlogN***, so the multiplication we need to take is ***0.5NlogN***. [4]

For an N-square-sized 2D FFT, we need $\boldsymbol{0.5N^2 logN^2 = N^2 logN}$ multiplications.

For an N-cube-sized 3D FFT, we need $\boldsymbol{0.5N^3 logN^3 = 1.5N^3 logN}$ multiplications.

In this analysis the size of the detector is set as ***N by N***, the size of the 3D image to be projected is N-cube-sized, the number of projection angles is set as $\boldsymbol{\pi N}$. This is an acceptable settings because in practice we usually take much more views than ***N***.

### 2.4.1 The complexity analysis for the NUFFT projector

The complexity analysis in this section is made under the conditions of the theoretical conclusion of the sampling rates on **(ρ, θ, φ)**, not the sampling rates we actually taken in the experiments. Recall the theoretical results, for a **N**-cube sized 3D image, the $N\rho=N$, $N\theta=\pi N$ and $N\varphi=\pi N$. And for an N by N sized projection data matrix, the $N\rho=N$ and $N\theta=\pi N$.

**Complexity for step 1:**

In step 1 of the NUFFT projector, we need a 2-time oversampled NUFFT, and the interpolation neighbour is chosen to be [3 3 3], so the number of the neighbours is 27, the number of the interpolation's targets is $\pi^2 N^3$ according to the sampling rates setting, the FFT size in φ and θ direction should be the nearest power of 2 larger than π, which is 4, so the number of multiplication we must do in the 3D NUFFT is:

$$C1a = 0.5 * 4 * 4N^3 \log(4 * 4N^3) + N^3 + 27\pi^2 N^3$$

$$C1a = 24N^3 \log N + 299.47N^3 \quad (2.16)$$

The 3D derivative:

$$C1b = \pi^2 N^3 = 9.87N^3 \quad (2.17)$$

The radial IFFTs:

$$C1c = (\pi N)^2 * 0.5 * N \log N = 4.93N^3 \log N \quad (2.18)$$

**Complexity for step 2:**

The step 2 is the 3D resampling step. In section 2.1.3, there are two proposed resampling method, and here is the complexity analysis for both of them:

For the 1st resampling method suggested by formula 2.11 a 3D IFFT and a 3D NUFFT are applied:

$$C21 = 0.5 * 16N^3 \log(16N^3) + N^3 + 0.5 * 16N^3 \log(16N^3) + 27\pi^2 N^3$$

$$C21 = 48N^3 \log N + 340.34N^3 \quad (2.19)$$

The second resampling method suggested by formula (2.15) is rather simple in complexity. Usually the sparse matrix 2 is chosen to be a reordering matrix (only have 1 element each row). The interpolation neighbour is chosen to be [3 3 3], a fair comparison with the first method. So the computation cost of the second method is only:

$$C22 = 27\pi^2 N^3 = 266.48N^3 \quad (2.20)$$

**Complexity for step 3:**

The third step for forward projector is the Grangeat step.

The cost for the de-postweighting is:

$$C3a = \pi^2 N^3 = 9.87N^3 \quad (2.21)$$

The radial FFTs:

$$C3b = (\pi N)^2 * 0.5 * N \log N = 4.93N^3 \log N \quad (2.22)$$

The 2D NUFFT transpose, the interpolation neighbour is set as [5 5]:

$$C3c = (\pi N) * ((2N)^2 \log(2N) + (2N)^2 + 25(2N)^2)$$

$$C3c = 12.56N^3 \log N + 339.28N^3 \quad (2.23)$$

The cost for de-preweighting is:

$$C3d = (\pi N) * N^2 = \pi N^3 \quad (2.24)$$

**Total complexity**:

After summarize the analysis of the complexity of each step, we can conclude that the total complexity of the NUFFT projector with the first resampling method is:

$$C1(total) = 94.42N^3 \log N + 1001.98N^3 \quad (2.25)$$

The complexity of the NUFFT projector with the second resampling method is:

$$C2(total) = 46.42N^3 \log N + 928.12N^3 \quad (2.26)$$

### 2.4.2 The estimation of the complexity of a CT projector

This section is about giving a complexity estimation of the simplest version of the linear CT projector.

The core principle of the CT projector is that: in each projection, at first reverse the cone beam geometry to parallel beam by interpolate the object to a parallel grid, then sum each slice up, as shown in the following figure:

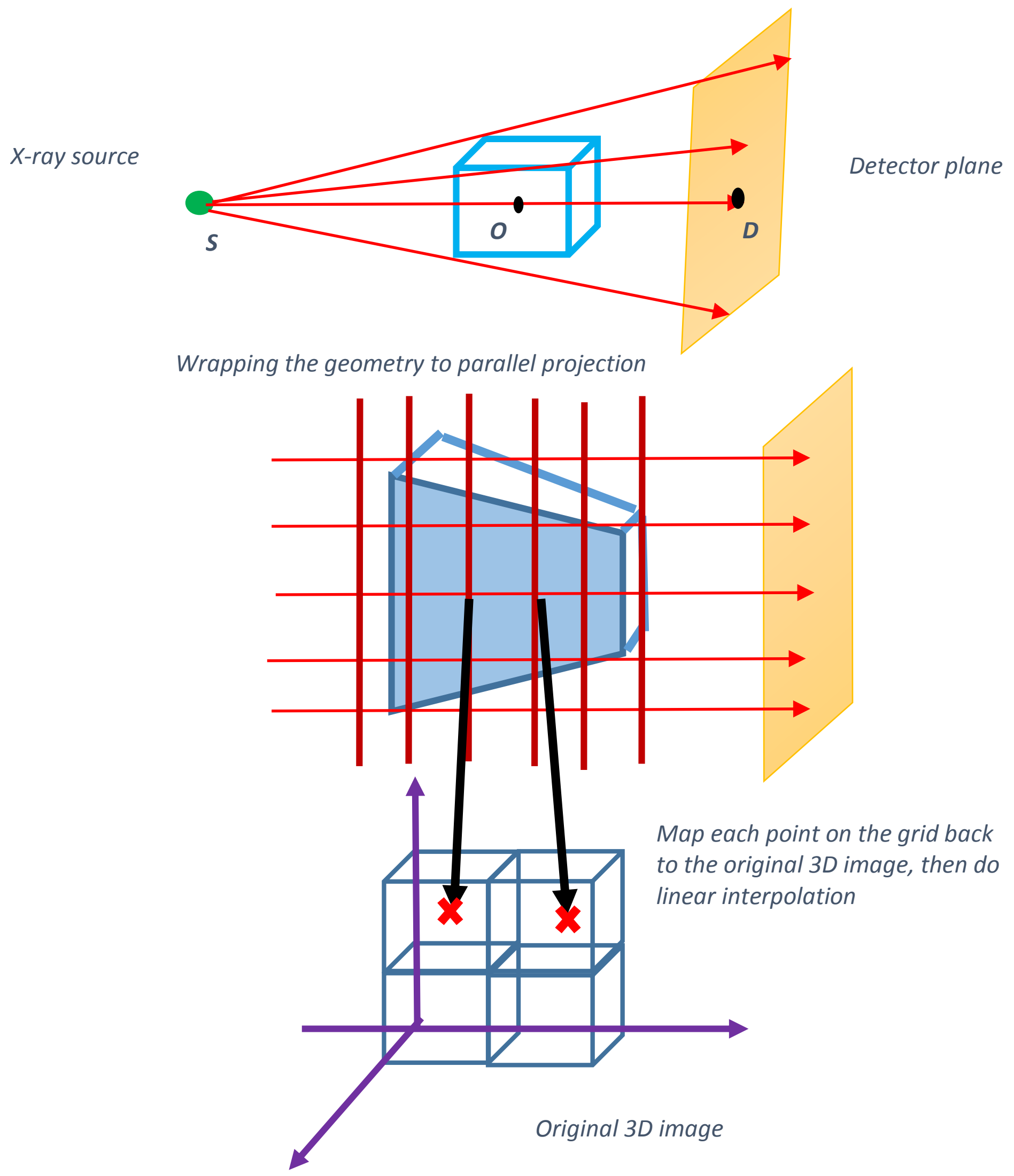

***Figure 2.17 The principle of the CT projector: interpolate the object to parallel projections***

It is clear that the dominant part of the computation is the linear interpolation from the 8 nearest points from the input 3D image.

If we use the same system's size described in section 2.4.1, as a fair comparison with the NUFFT projector: the size of the input image is N cube, the size of the detector matrix is N by N, and the number of the projections we take is also the nearest interger of $\pi N$, the complexity of the CT projector is:

$$C3(total) = 8N^3 * \pi N = 25.13N^4 \tag{2.27}$$

Note that this is the most simple CT projector one can get, because it use the minimum number of interpolation neighbours (which is 8 in 3D case), and there is no oversampling involved at all.

### 2.4.3 The complexity comparison between the NUFFT projector and the CT projector

The complexity expression formulas for each projector, (2.25), (2.26) and (2.27), have common factor $N^3$. Because both three functions will return huge values when N is large, it is better to divide them by this common factor $N^3$ before doing comparison. And the following figure shows the comparison between the three projectors:

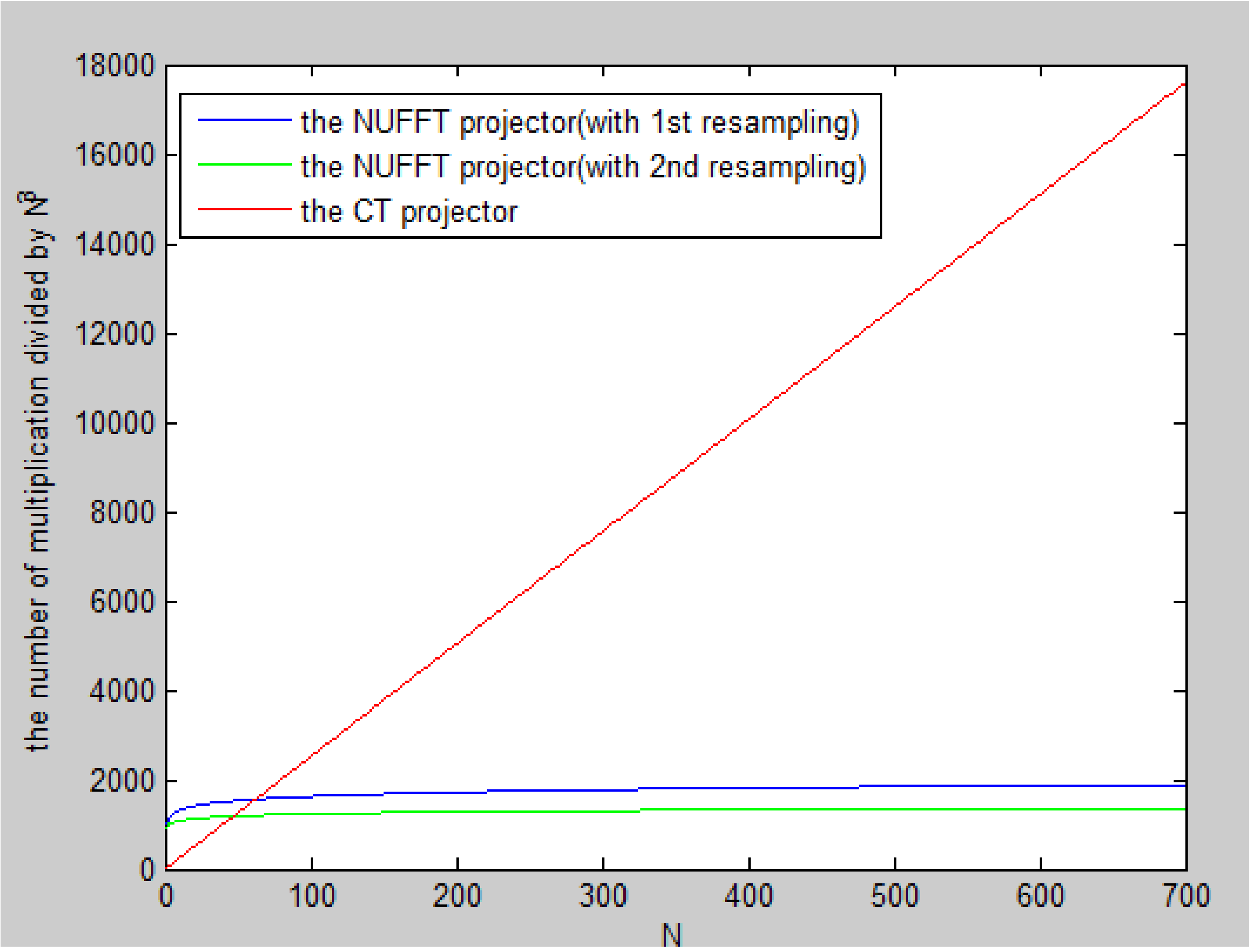

*Figure 2.18 The comparison of complexity between the NUFFT projectors and the CT projector*

From the result of the complexity analysis, we can see that the complexity of the NUFFT projector will have dominant advantage in terms of complexity over the simplest version of the CT projector when N is large. For example, when N=512, (512-cube size is the practical size of a 3D CT image nowadays which can suit the clinical need), the NUFFT projector with the 1st resampling method will be about 7 folds faster than the CT projector, while the NUFFT projector with the 2nd resampling method will be about 10 folds faster than the CT projector.

Note that this complexity analysis is strictly following the theoretical minimum angular sampling rates given by chapter 1, and the designs of the NUFFT forward projection algorithm, so its result is reliable and able to justify the advantage of the NUFFT forward projector in terms of complexity, although the projection algorithms implemented in the experiment part of this report do not consistent with the result because of the different sampling rate actually chosen to be used.

# 3. The implementation of the algorithms and experiments

This chapter is for the implementing and testing of the algorithms which has been designed and check their performance.

## 3.1 Experiment for determining the radial line sampling number in practice

The first experiment is the testing of the information-preserving ability of angular line sampling pattern and determine a practical number of radial line sampling for the application of the Fourier Slice Theorem based algorithms designed in this project.

In chapter 1 we have already discussed about the minimum equal spaced line sampling from the theoretical perspective. But the theoretical result is not used in this experiment because it use different sampling rate to (θ, φ, ρ) and make the NUFFT projectors more complicated to be implemented in the limited time of the project. So **in this project's implementation and experiments of the NUFFT projector, we are aimed to make a convenient and quick research to see the practicality and rough performance of the NUFFT projectors designed in the previous chapter.** In chapter 1 we have derived the theoretical bounds for the sampling rates on the **(ρ, θ, φ)**, and the method of reducing a half of the sampling rate on θ by up-sampling ρ by 2 and still preserve all the information of the 2D Sinogram, and the experiment 1 is aimed at numerically justify this trade-off.

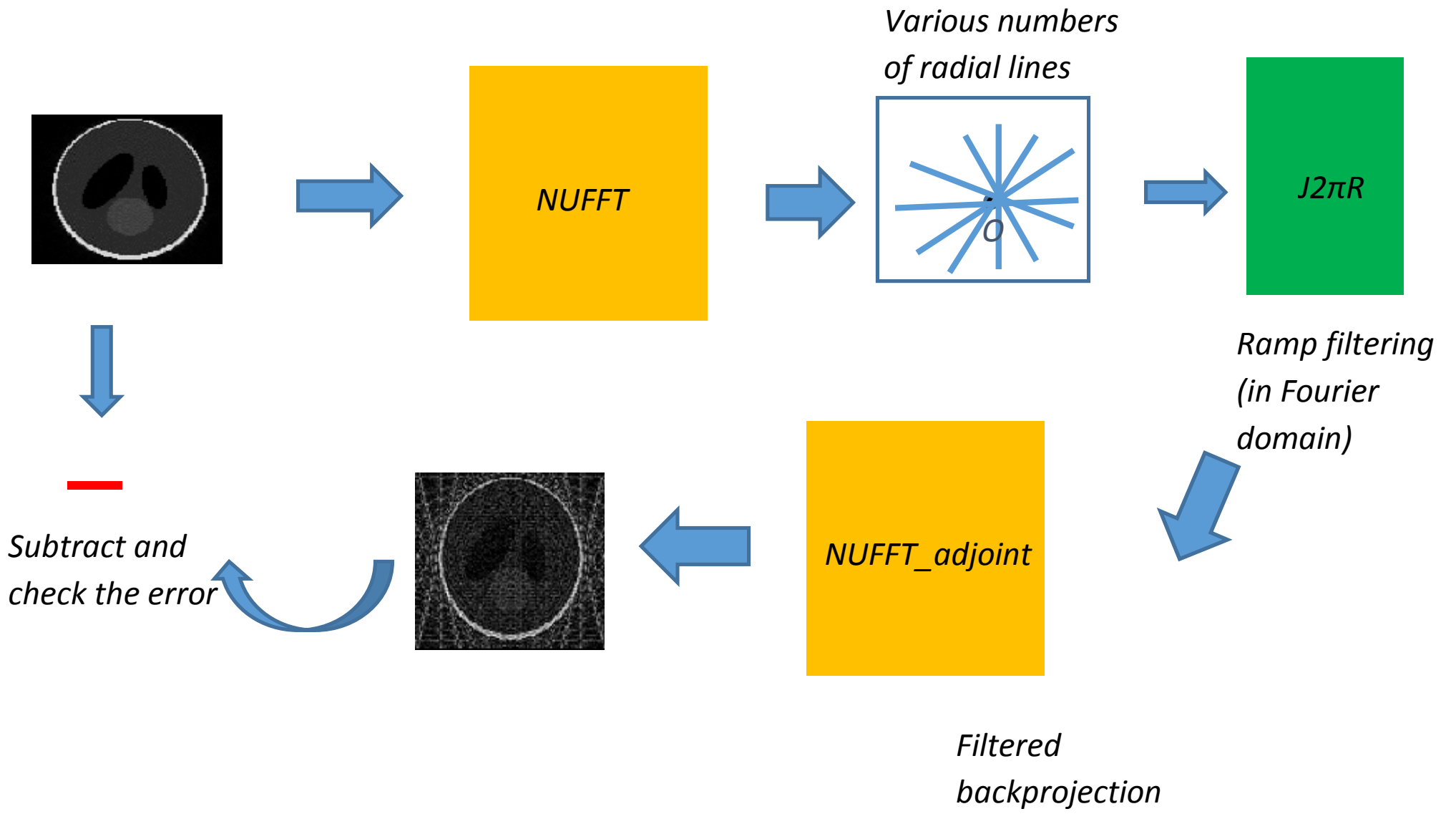

***Figure 3.1 the structure of the experiment 1***

In this section, an experiment aims to numerically determine the angular sampling number for a certain sized image will be designed and executed.

The plan of the experiment is shown in figure 3.1.

The idea for the experiment is to:

1) Take a standard and well acknowledged medical image (for example the Shepp-Logan phantom) as the object for the experiment. In this experiment the image size is choose to be 128 by 128.
2) Use the NUFFT to project the image into radially sampled Fourier space, the angular sampling rate (the sampling against θ) increase from 16 to 384, and the sampling rate along each line (the sampling against ρ) is fixed at 256. This sampling rate for ρ is consistent to the theoretical result given by chapter 1, which shall be twice as many as the size of image (in this case, is 128). Because of the Fourier Slice Theorem, one can easily see that taking radial sampling in the Fourier domain is equivalent to taking the same pattern of radial sampling in Radon space (the Sinogram).
3) Apply filtered backprojection from the radially sampled Fourier space by ramp-filtering and NUFFT transpose.
4) Subtract the original image and the reconstructed image, calculate the average error per pixel.
5) Increase the number of angular sampling lines and repeat from step 1. And then see at which point the error stops to reduce.

The first image we tested is the $60^{th}$ slice of the 3D Shepp-Logan phantom (128 cube sized). And the following figure shows the reconstructed image from different number of angular line samplings, from these images we can observe that by increasing the angular sampling the error of the reconstructed image will be reduced.

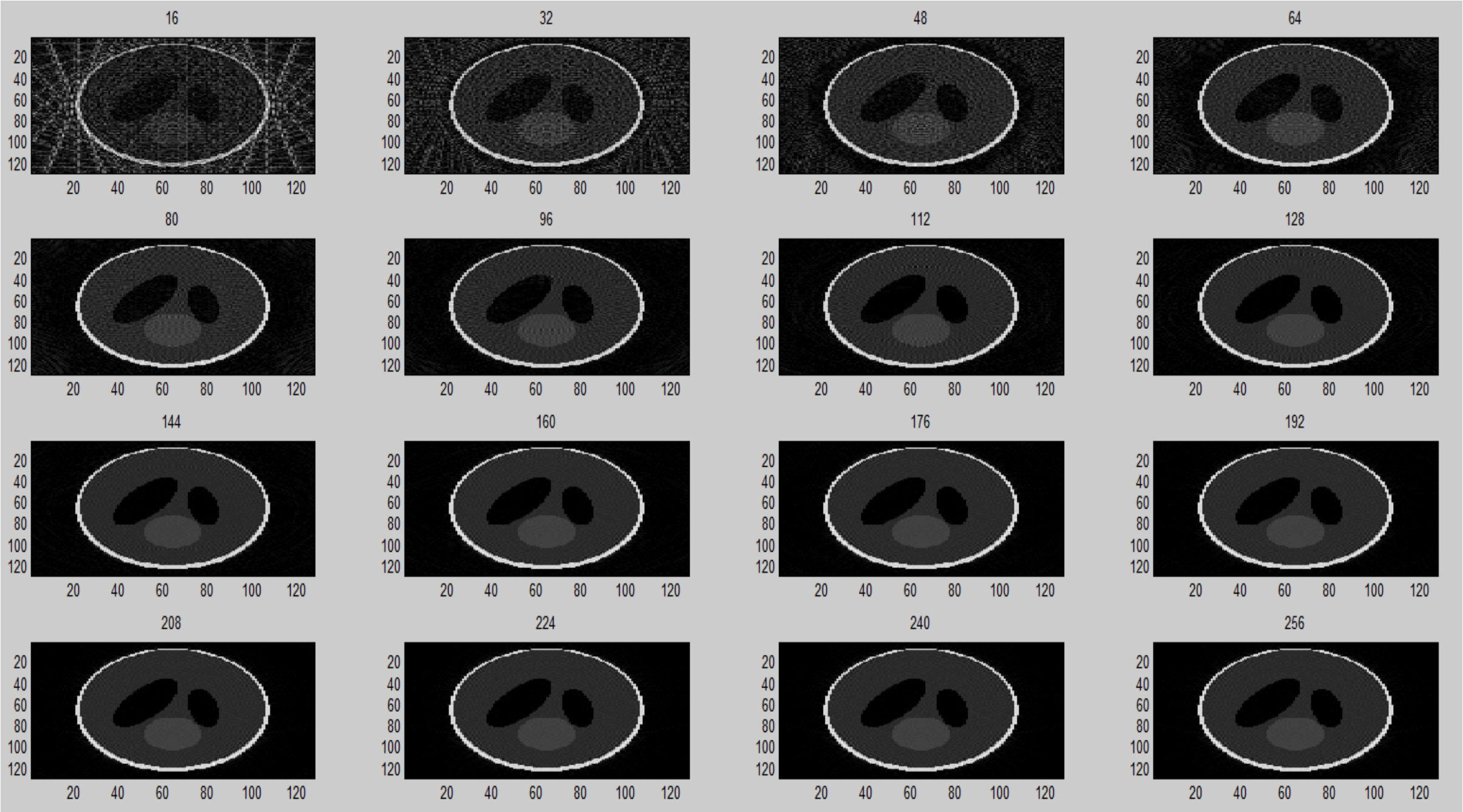

***Figure 3.2 the reconstructed images from different number of angular samplings***

And let us look at the difference image from the original image:

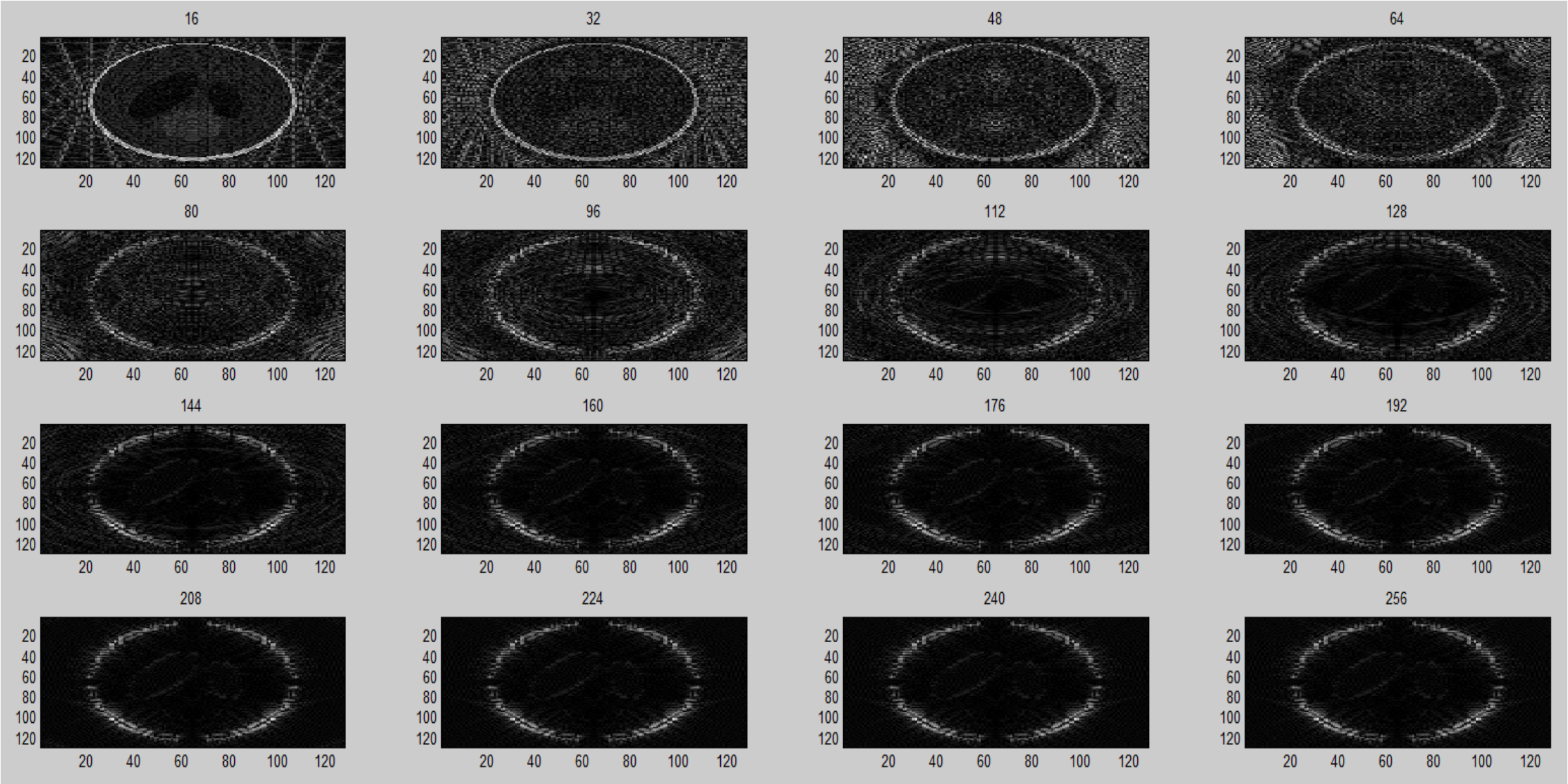

***Figure 3.3 the difference images between the reconstructed images and the original images***

And the error plot:

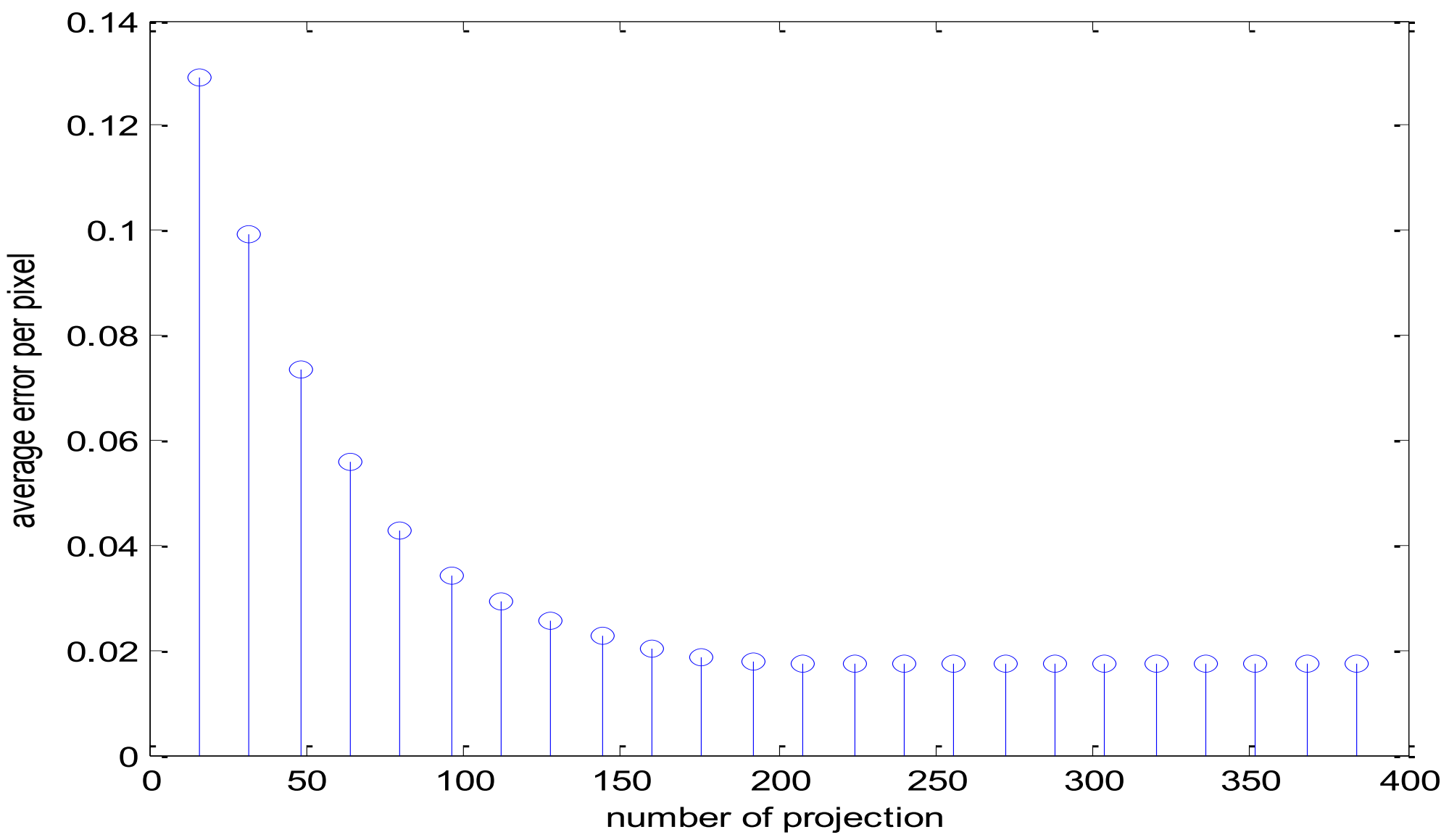

***Figure 3.4 The error plot of the reconstructed image***

We can see that for a 128 by128 sized image the average error per pixel stopped to reduce after the 200 radial line sampling, which is much lower than the theoretical result. To make this more convincing, some more images are also chosen to be tested. Let us look at the result for the slice 80 of 3D phantom, an image for real human body slice, and a random image also:

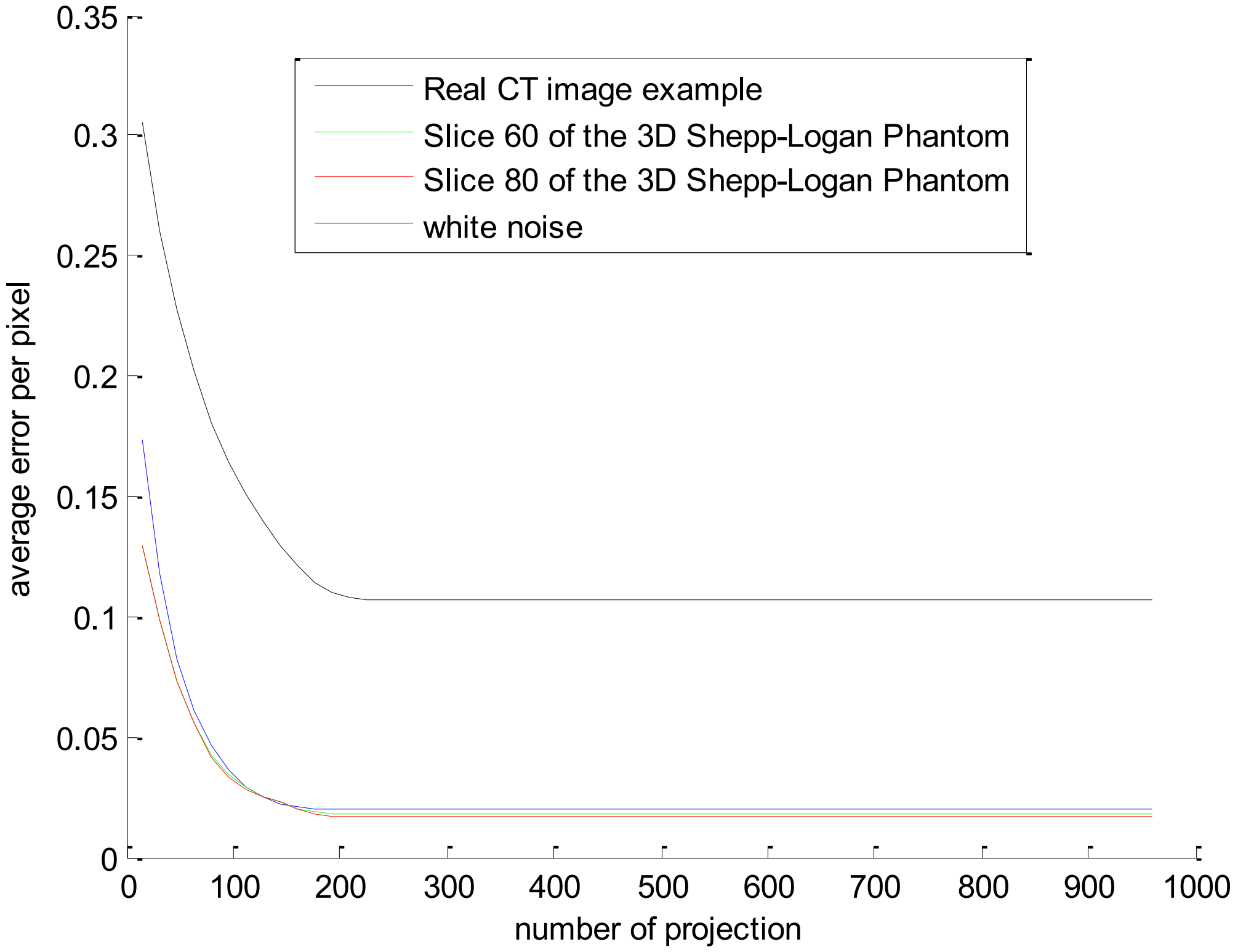

*Figure 3.5 The error plot of the images chosen to be tested*

The reason of the existence of the constant error which cannot be quenched by increasing the angular sampling number, as shown in figure 3.5, will be mentioned in chapter 4. It is basically due to an inherent deficiency of the radial sampling pattern.

From the results of the experiments on these 4 images, we can see that for a ***128 by 128*** image, the minimum number of angular sampling in Fourier space (and Radon space) is approximately ***200*** when we sample ρ by ***256*** on each radial line. Note that ***128*π/2*** is approximately ***200***, so this numerical result confirm the theoretical conclusion in chapter 1. Because of this observation, we can numerically justify the choice of the angular sampling on ρ and θ in the algorithms designed in this project:

For an ***N by N*** image or data set, the ***2N*** samples will be taken on ρ and θ.

But for an ***N cube*** sized 3D image, ***2N*** angular samples will be taken against ρ, the azimuth angle φ and elevation angle θ as well. The sampling number on φ is still not adequate according to the theoretical derivation in chapter 1, and this is the compromise point of the experiments recorded in this report.

## 3.2 Experiments on the NUFFT based forward projection algorithm

In section 2.1, the derivation of the 3D resampling method, formula (2.11) and formula (2.15) represents two different methods of performing this resampling: the first one is NUFFT based, and the second one only utilize the interpolation operator of the NUFFT. In this section the forward algorithm with both ways of resampling will be tested. In this project the NUFFT algorithm and its transpose algorithm's implementation is given by J. Fessler's Image Reconstruction Toolbox [13].

Note that because of some difficulties in implementing the proposed resampling method efficiently in Matlab in limited time, in this experiment, for the NUFFT projector with the resampling method given by formula (2.11), an NUFFT-transpose is used to replace the Inverse Fourier Transform. Formula (2.13) has justified the replacement for us because the input data is uniformly located.

But because the NUFFT-transpose is used to replace the IFFT, the forward projector's complexity achieved in this project is not yet consistent to the result given by section 2.4.

The implementation of the $2^{nd}$ resampling method has the similar problem. Ideally the $2^{nd}$ sparse matrix in formula (2.15) should be simply a reordering matrix (a matrix which has only one element each row). But the NUFFT toolbox given by J. Fessler cannot generate such a reordering matrix, the sparse (identity) matrix 2 is replaced by an interpolation matrix with [3 3 3] interpolation neighbour, and a 2 time oversampling is used.

### 3.2.1 The geometric settings for forward algorithm test

Now the cone beam projection model and its settings is given below, shown by figure 3.11.

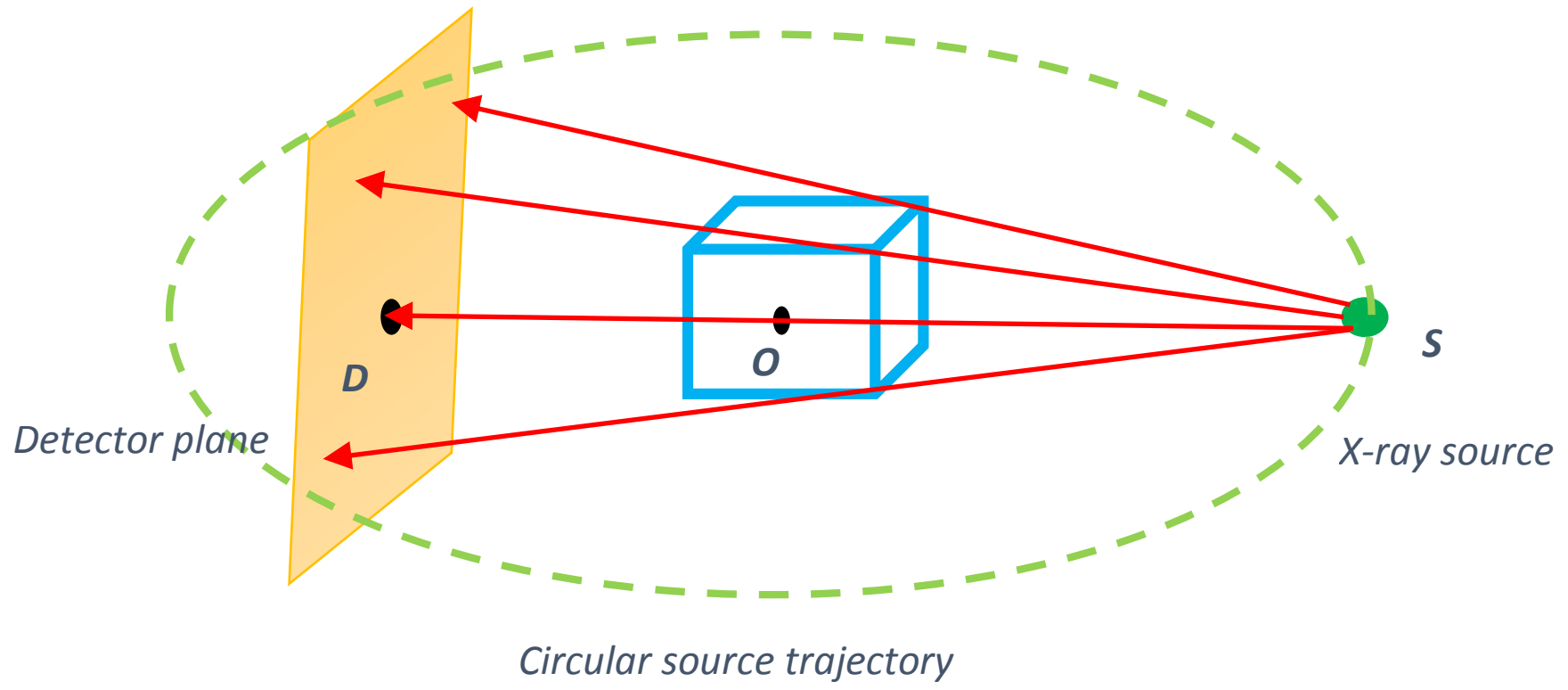

***Figure 3.6 The geometry of cone beam CT, where the point S represent the X-ray source, point O is the central point of the object, and D is the central point of the detector plane***

As shown in figure 3.11, the cone beam projection model is a standard one: the X-ray source, the central point of the object and the central point of the detector plane are both in the same line, and meanwhile in one projection the whole volume of the object is covered by the range of the X-ray beam.

The distance between the X-ray source and the central of the object (SO) is set as 1100 mm, and the distance between the X-ray source and the central point of the detector plane is set as 1500 mm. The detector's size is 512 mm * 512 mm, and the size of the object is set as 360 mm *360 mm *360 mm.

Now is the second part of the parameter setting, it is for determine the size of the algorithm. For 128 cube size algorithm, the number of detectors in the detector plane is 128 by 128 so the detector's data matrix is sized 128 by 128, and the number of projection along the circular source trajectory is 256, and the X-ray source's emission positions are also equally spaced.

### 3.2.2 The results of forward algorithm and the comparison with the CT projector

The CT projection operator from Kyungsang's Matlab toolbox [12] is chosen as a comparison in this section. The fundamental principle of this operator is linear interpolation and summation, and is typically slow, but the accuracy is reliable since the core principle is consistent with the physic-nature of X-ray CT projection. For this reason, this CT projector is a good choice for testing the NUFFT based forward projector and check whether it is able to produce projection data which is close enough to the real projection data.

The experiment is run at the geometry described in 3.2.1. The following figure is the cone beam projection data of a 128 cube-sized 3D image, generated by NUFFT based projector with the first resampling method proposed in chapter 2, and is scaled down by a constant to normalize the pixel values in the projection data set. Then the next figure is the cone beam projection data sets generated by the CT projector. And then follows the difference image between the two sets of results.

From the figures of these results we can see that the NUFFT forward projector designed in this project can have very good accuracy inside the phantom when referenced to the projection data given by CT projector. But currently it has some problems to be investigated and fixed in the future: The NUFFT forward projection algorithm generates an error at the region outside the object, and at some of the projection data set, there are some erroneous stripes. The root cause of these errors are most likely due to some bugs in the NUFFT's toolbox, because when we test the NUFFT projector with the second resampling method with the same settings, the stripes become much fewer as shown in figure 3.11.

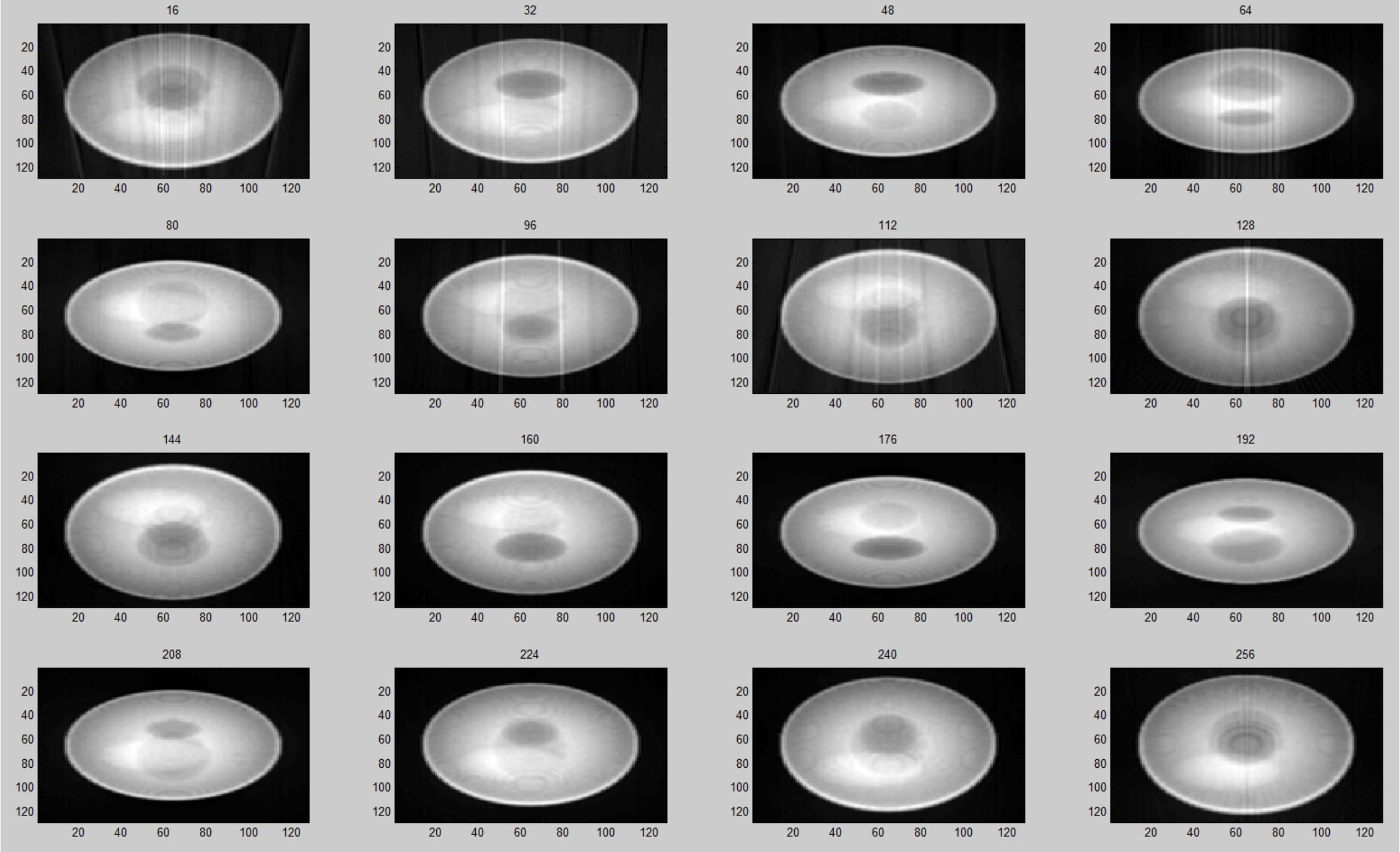

*Figure 3.7 the projection data generated by NUFFT projector with the first resampling method*

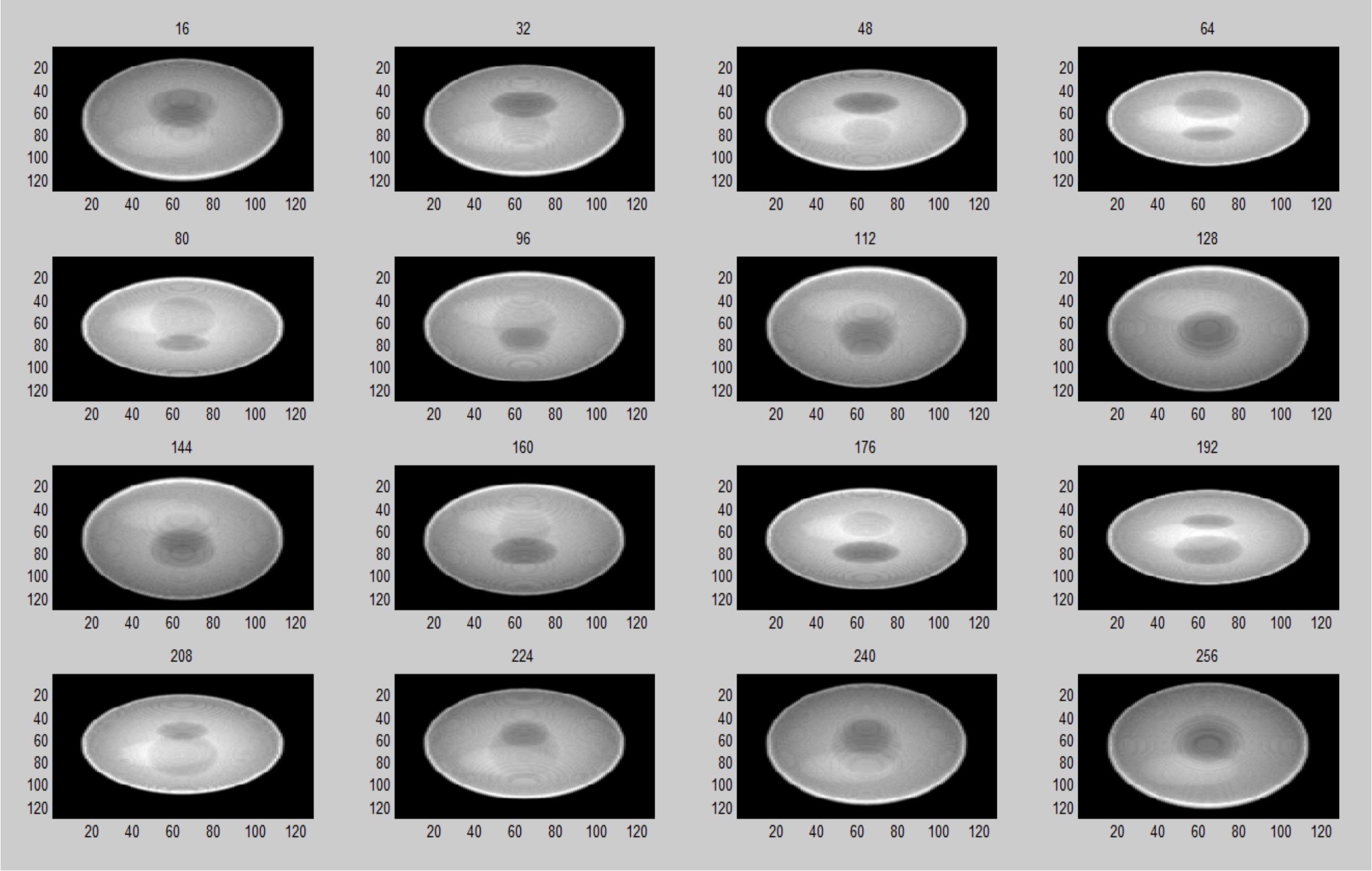

*Figure 3.8 the projection data generated by CT projector*

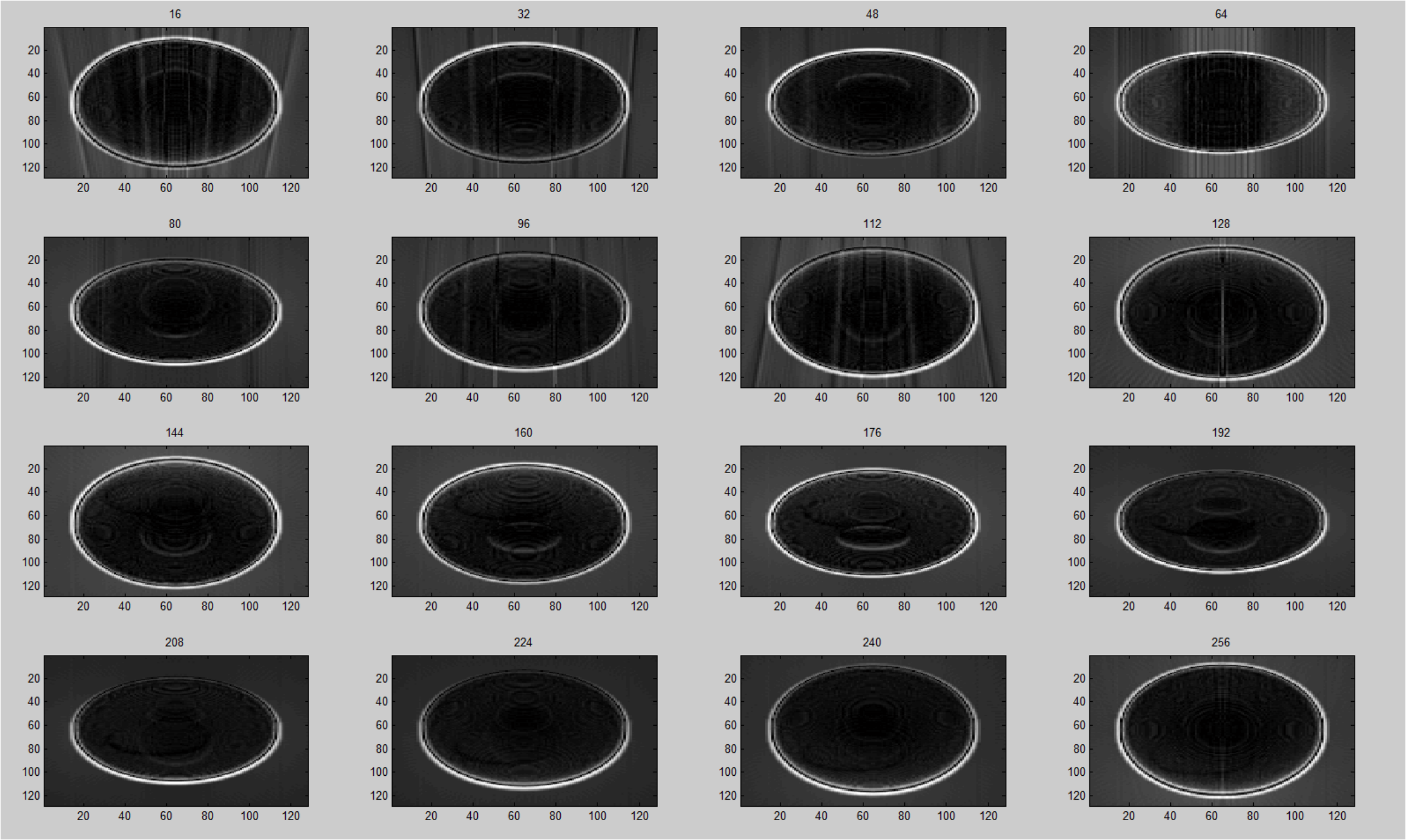

*Figure 3.9  the difference images between the projection data generated by the NUFFT projector and CT projector*

```
>> new_forward_algorithm_128
nufft_init 260: Needs at least 4.32e-07 Gbyte RAM
nufft_init 260: Needs at least 4.32e-07 Gbyte RAM
nufft_init 260: Needs at least 4.32e-07 Gbyte RAM

total_time =

   56.5960

>> tic;[ proj ] = CTprojection( x, param );t1=toc

t1 =

   78.6783
```

*Figure 3.10  The speed test of the two algorithms, it may not be a meaningful comparison since it can be affected by better implementations of these algorithms*

And then let us look at the results of the NUFFT projector with the second resampling method given in chapter 2. Because this method takes only the interpolation step of the NUFFT algorithm, without doing FFTs in the intermediate step, the projector will be more efficient in computation, and it takes about 44 seconds using the same implementation of Matlab code.  The stripes occurred in some of the former projection datasets is no longer there, as shown in figure 3.11.

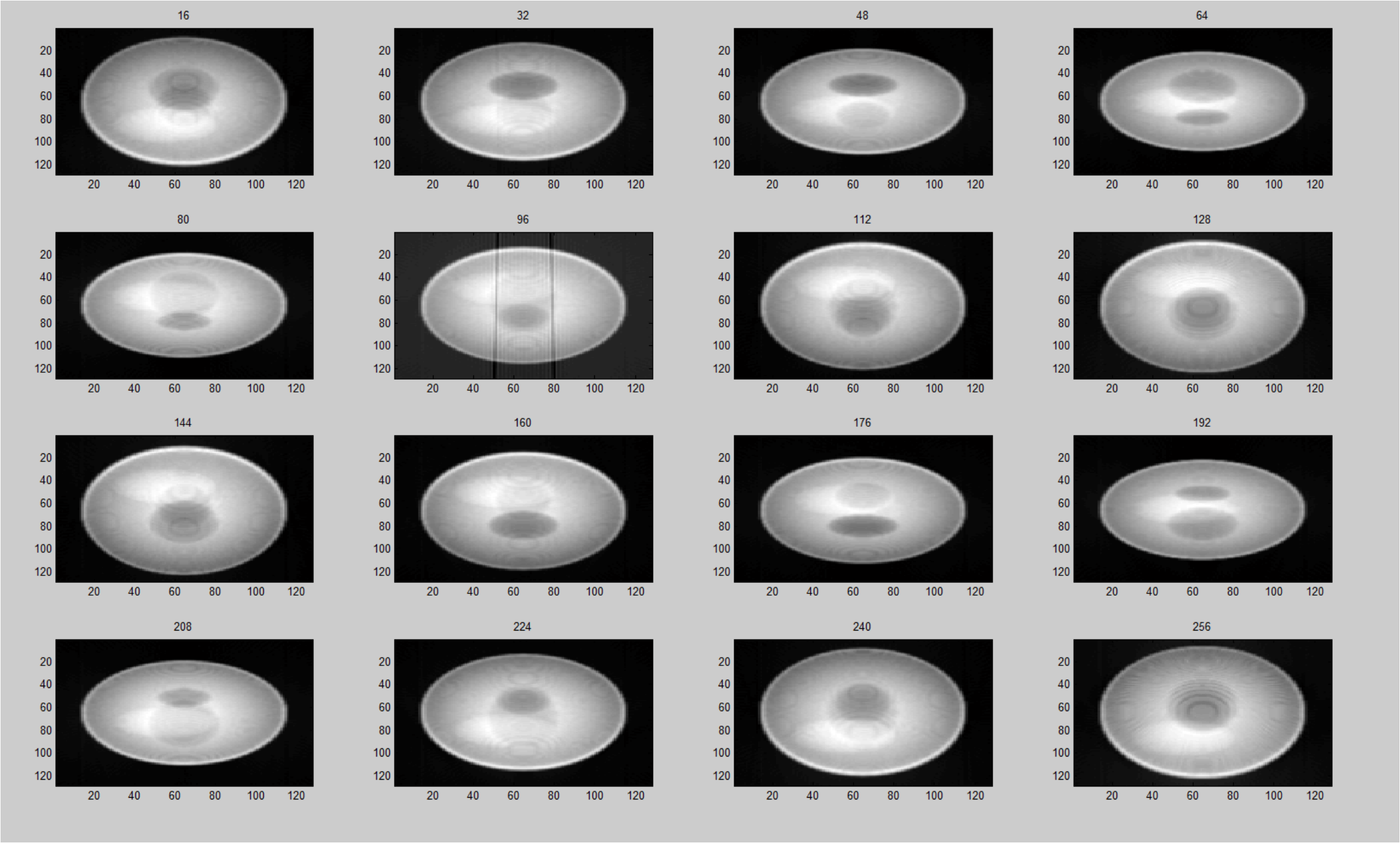

*Figure 3.11 The projection data generated by NUFFT projector with the second resampling method*

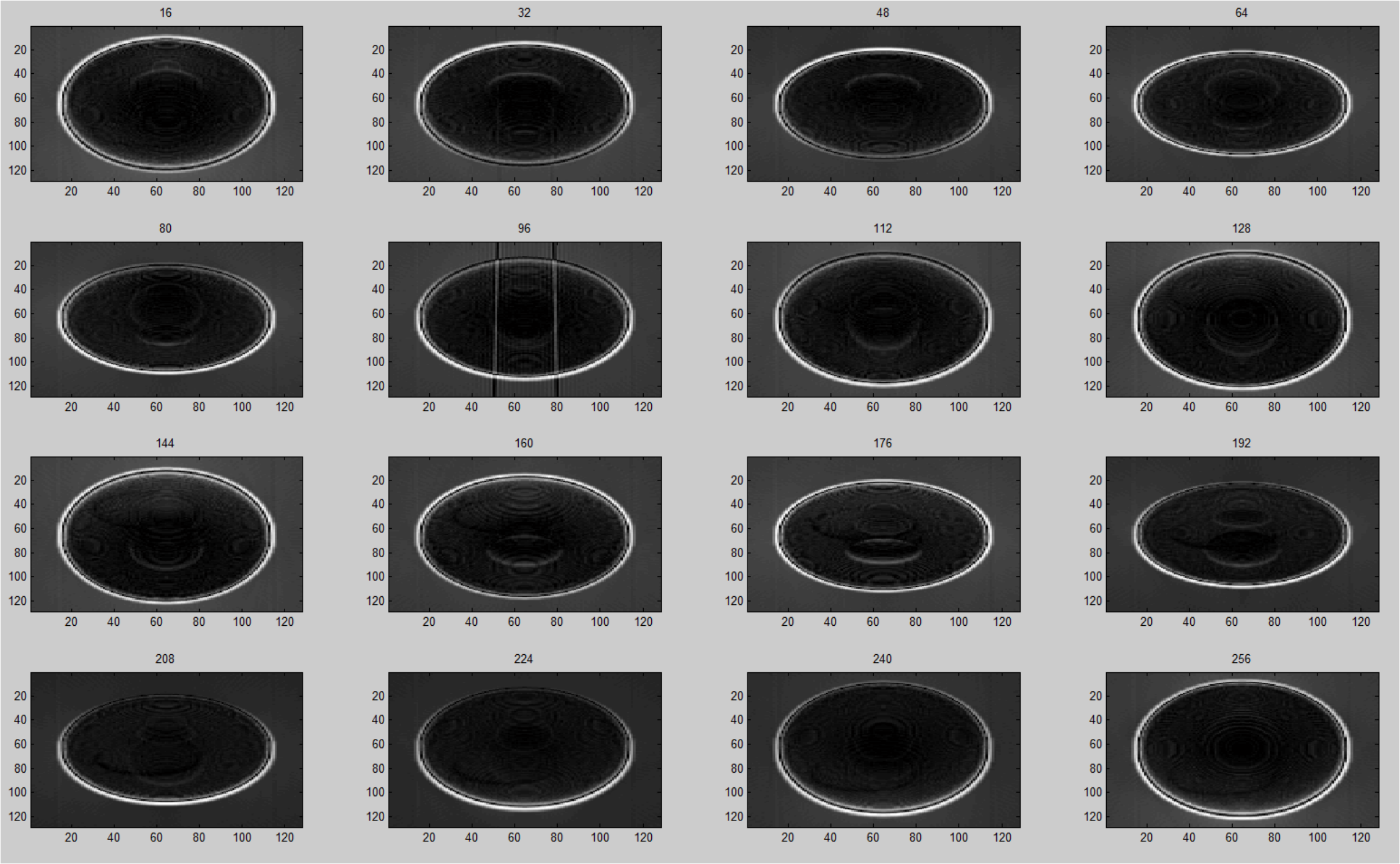

*Figure 3.12 The difference images between the projection datasets generated by the NUFFT projector with the second resampling method and the CT projector*

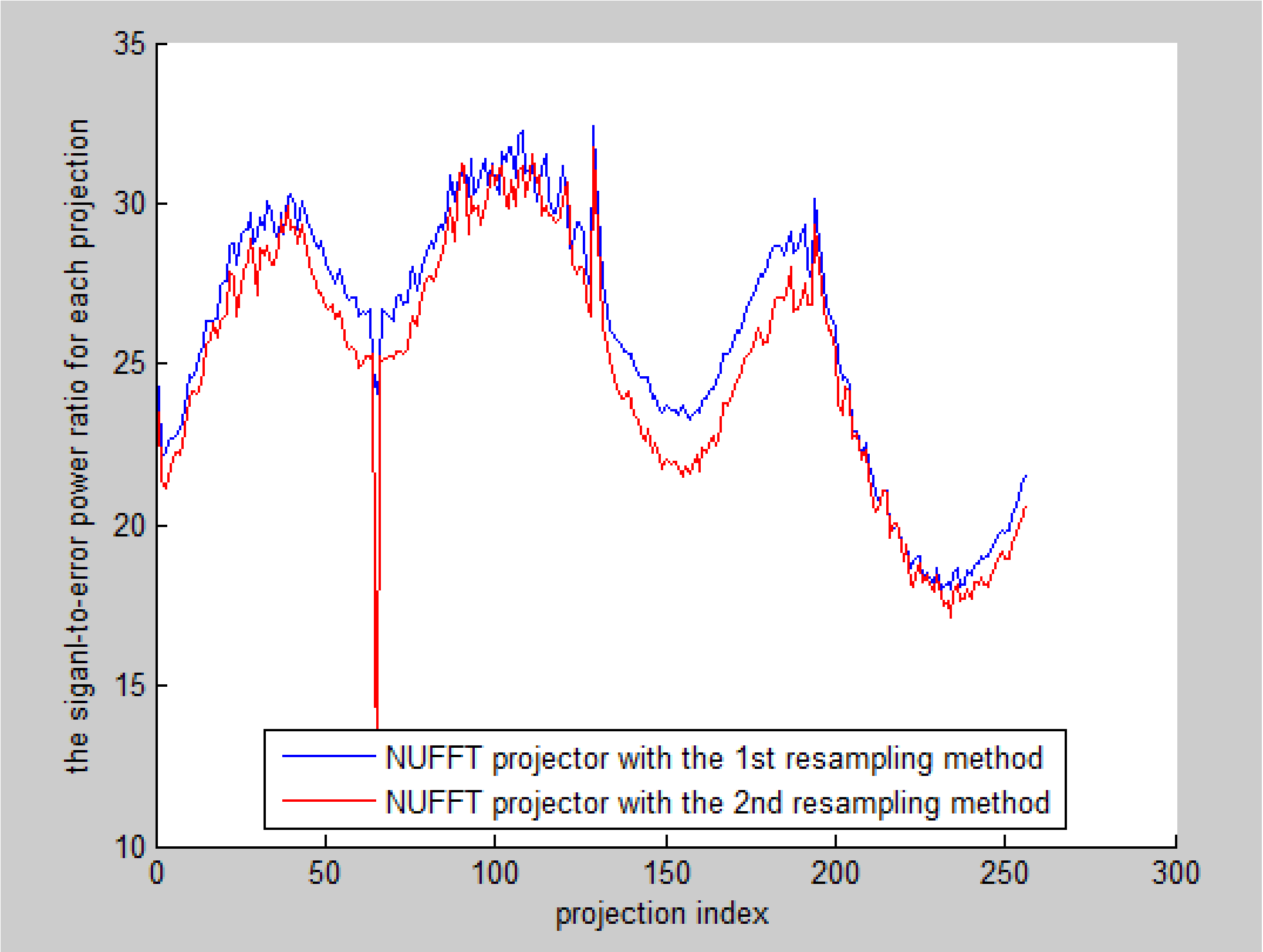

***Figure 3.13 the signal-to-"error" plot which is aimed to show the closeness between the two streams of the NUFFT projector and the CT projector. We can see that at projection 65 a serious inaccuracy happens. When we check the No.65 projection data set we found out that the stripes errors are concentrated at that particular set of data, and the reason is needed to be investigated in the future.***

The figure above is the accuracy comparison between the NUFFT projectors with different resampling method, assume that the projection data generated by CT projector is the ground truth. The measurement of the error for each sets of projection data generated by the NUFFT projector is the summation of the difference image and then average it to each pixel. Note that before doing this, we multiply the projection data set by a mask to null out the values outside the shell of the phantom, so we can make sure that we are only measuring the error which is at the shell or inside the phantom. From the quantified result we can see that the NUFFT projector with the 1$^{st}$ resampling method can perform slightly better than the one with the 2$^{nd}$ method in terms of accuracy.

Because according to the difference image, the error power is dominantly exist at the shell, and within the shell the projection data has only a fraction of the error, we conclude that the NUFFT based forward projector can provide a reliable approximation of the CT projector. Further convincing experiments which can demonstrate this conclusion will be taken in the next section when these projection data sets is used to reconstruct the 3D Shepp-Logan phantom.

## 3.3 The joint experiments of the backprojection algorithm and the forward projection algorithms

Strictly speaking, the backprojection algorithm designed in the chapter 2 is currently not a rigorous algorithm because the resampling method is initially designed for forward projection, and in this project's backprojection algorithm it is directly used in reverse.

But however, in spite of this trick, the other parts of the algorithm is correct. In fact, if the correct 3D resampling method for the backward direction is found and implemented, we can say that the backprojector is a completely data-consistent algorithm. Let us look at the performance of the backprojector through experiment.

This experiment tests the backprojector by choosing 1) the projection data generated by the NUFFT based forward projector with the second resampling method (because it gives better results in term of erroneous stripes shown in the figure 3.11), and 2) the projection data generated by the CT projector respectively as the input data sets. Note that since the forward projector generates constant error at the regions outside the shell of the phantom, we used a mask to null out these values before apply it to the backprojector.

The following figure is the reconstructed image from the projection data generated by NUFFT projector, and then is the difference image compared to the original 128 cube-sized 3D Shepp-Logan phantom.

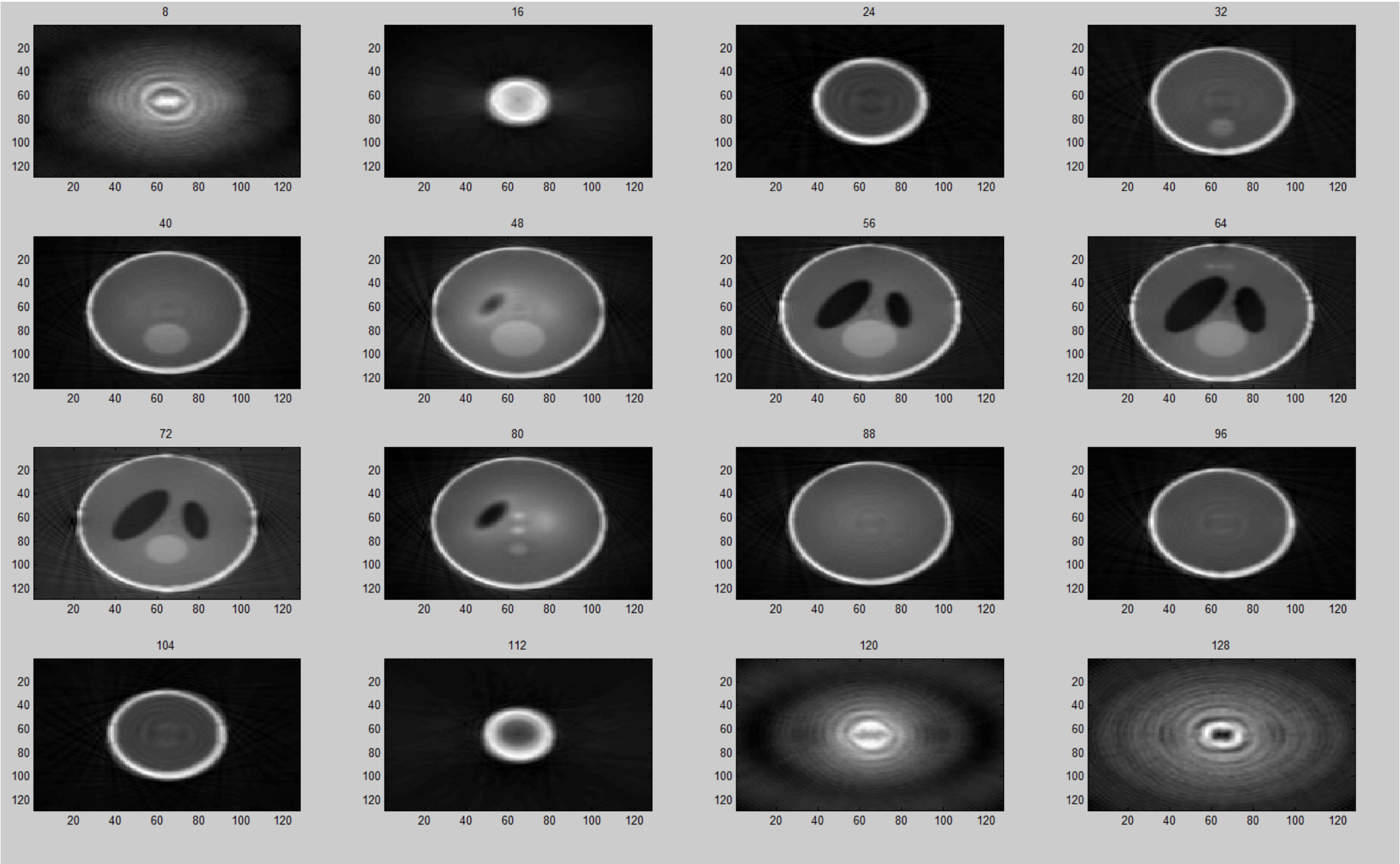

*Figure 3.14 The reconstructed image from the NUFFT forward projection's data*

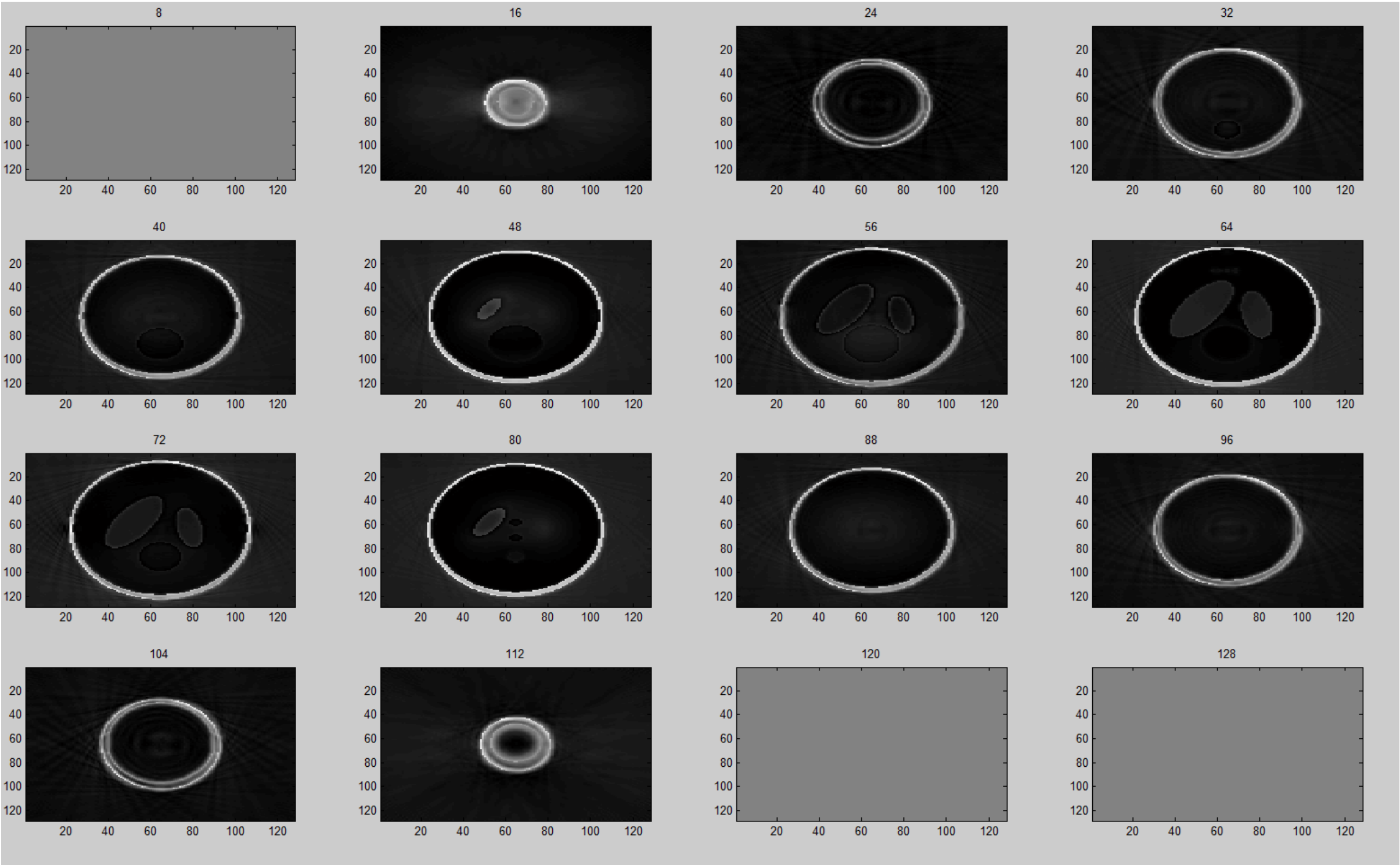

*Figure 3.15 The difference image between the results shown in figure 3.17 and the original Shepp-Logan phantom*

We can also test the NUFFT backprojector using the CT projector's data also:

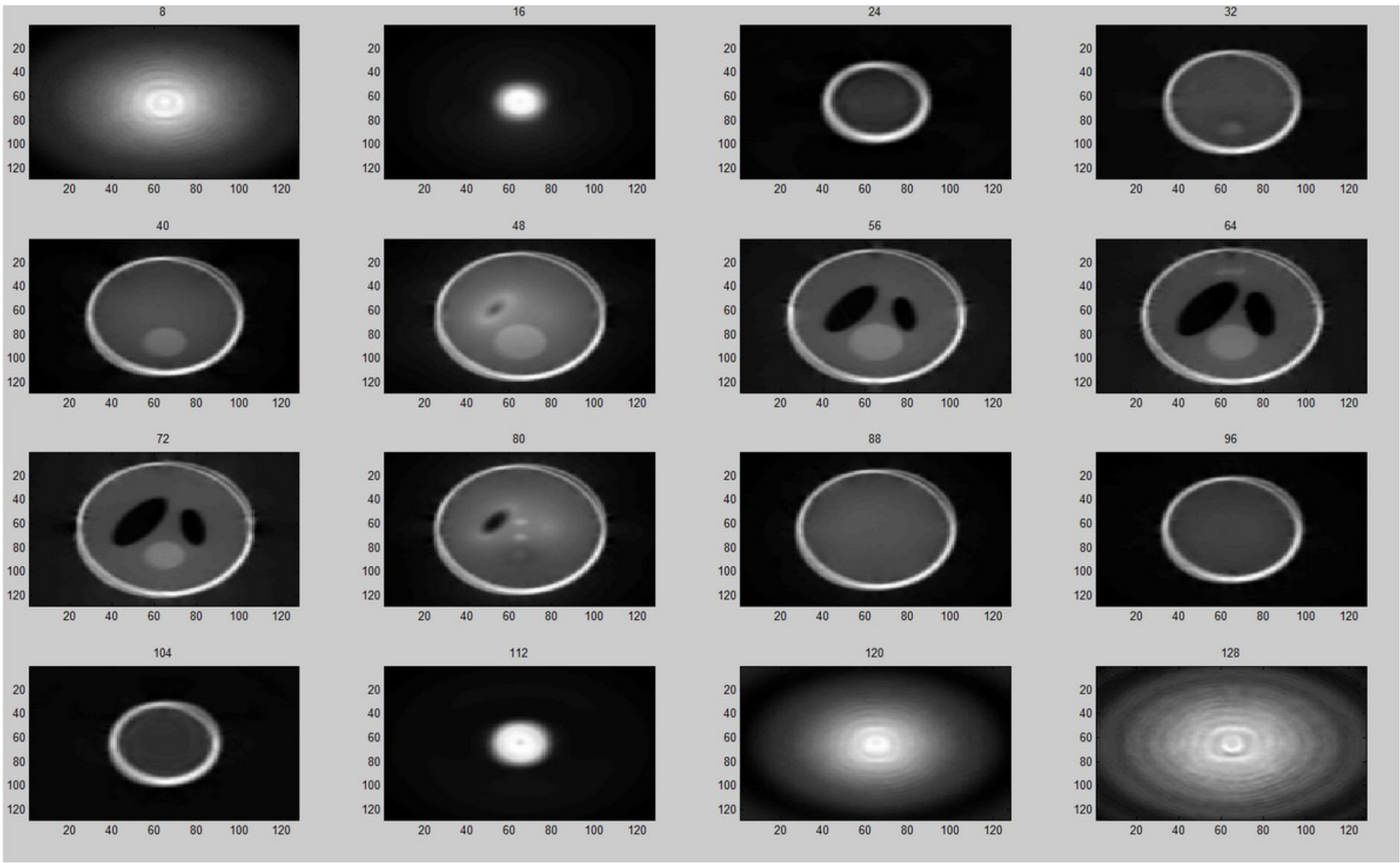

*Figure 3.16 The reconstructed image from the projection data given by the CT projector*

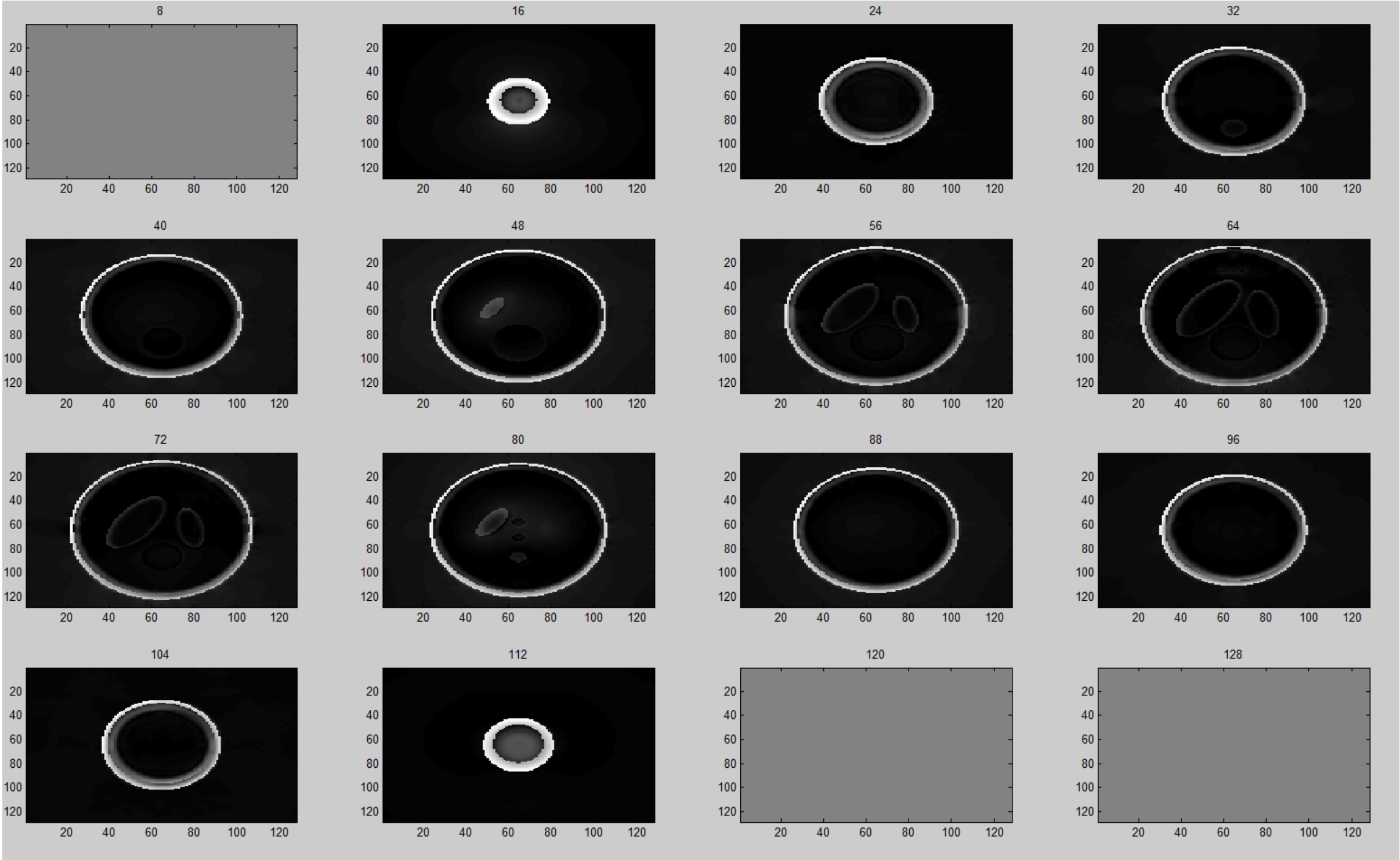

***Figure 3.17  The difference image between the reconstructed image in figure 3.19 and the original Shepp-Logan phantom***

For comparison, the result (difference image) of the reconstruction using the FDK algorithm is shown in the following figure:

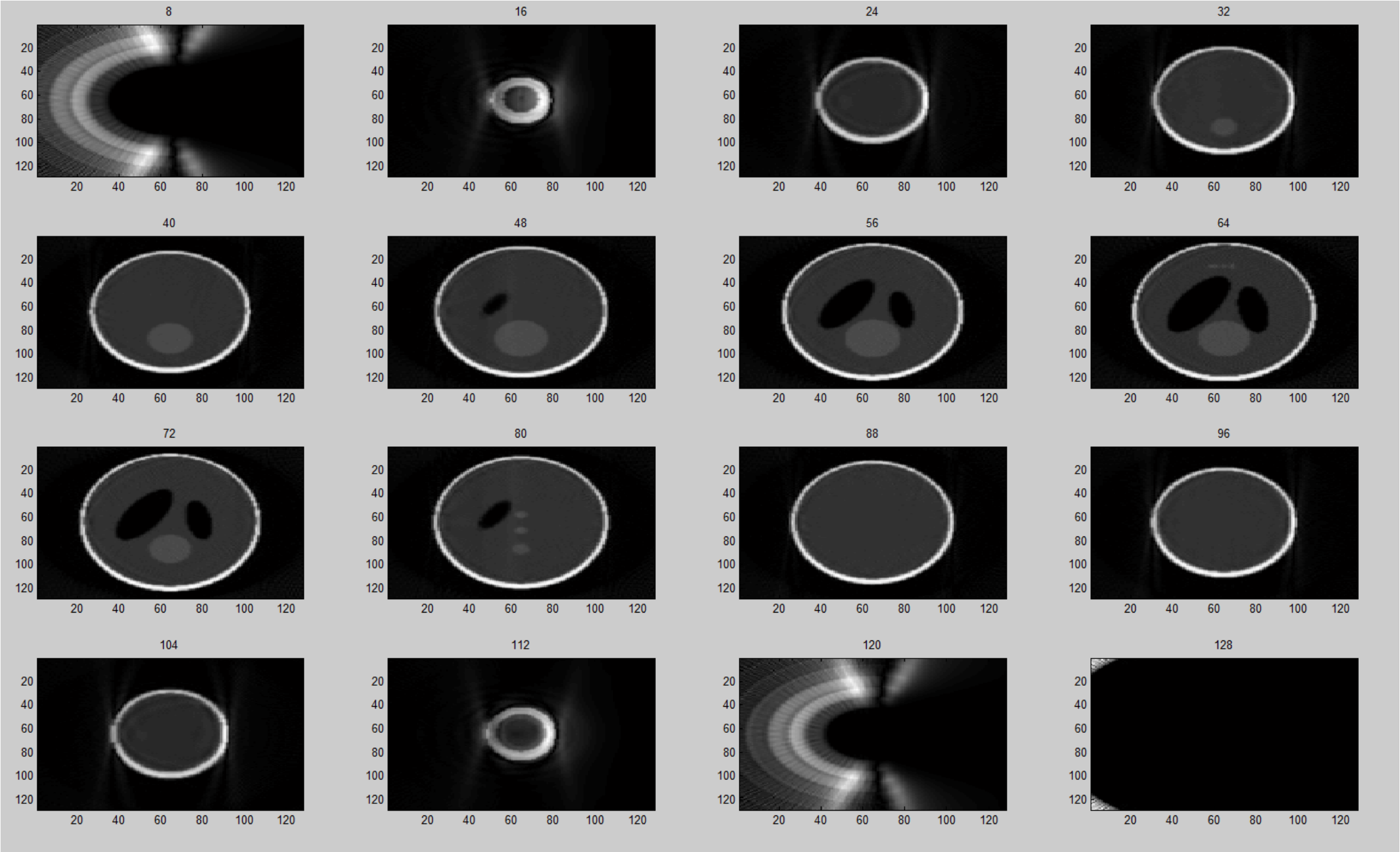

***Figure 3.18  The reconstructed Shepp-Logan phantom image by the FDK algorithm (from the projection data given by the CT projector)***

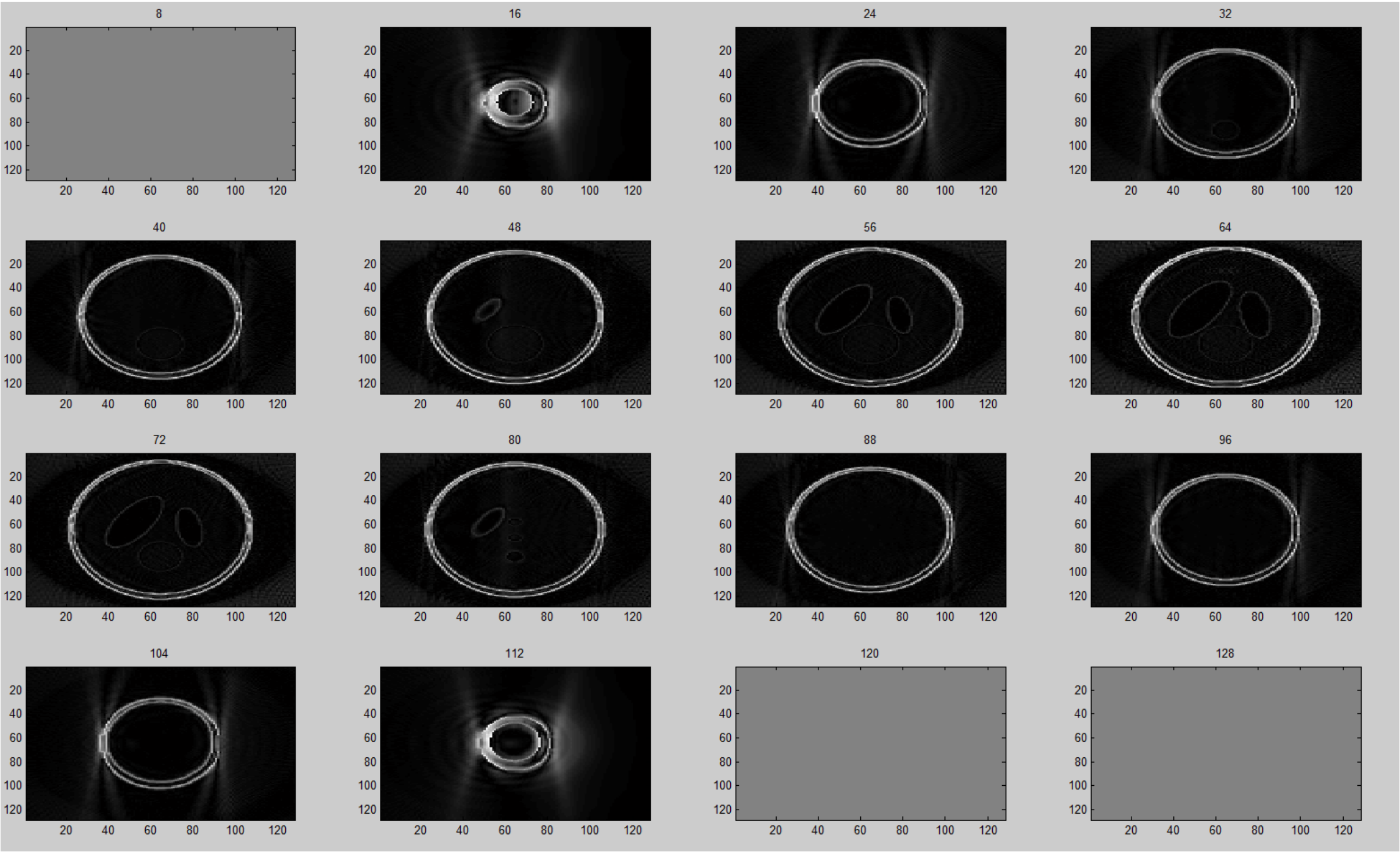

*Figure 3.19 The difference image between the reconstructed image by the FDK algorithm (from the projection data given by the CT projector) and the original Shepp-Logan phantom*

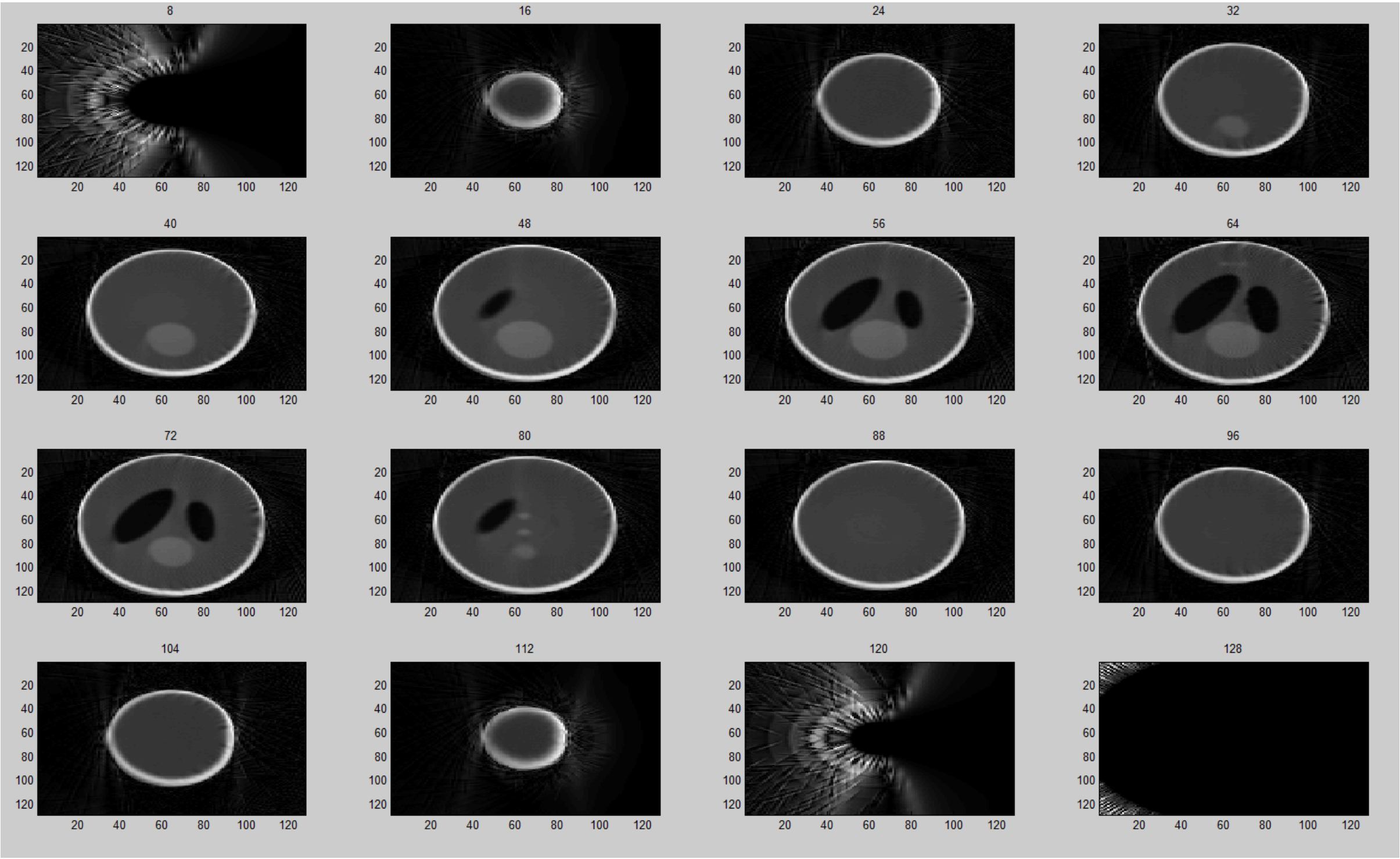

*Figure 3.20 The reconstructed images (reconstructed from NUFFT projector's projection data) by the FDK algorithm*

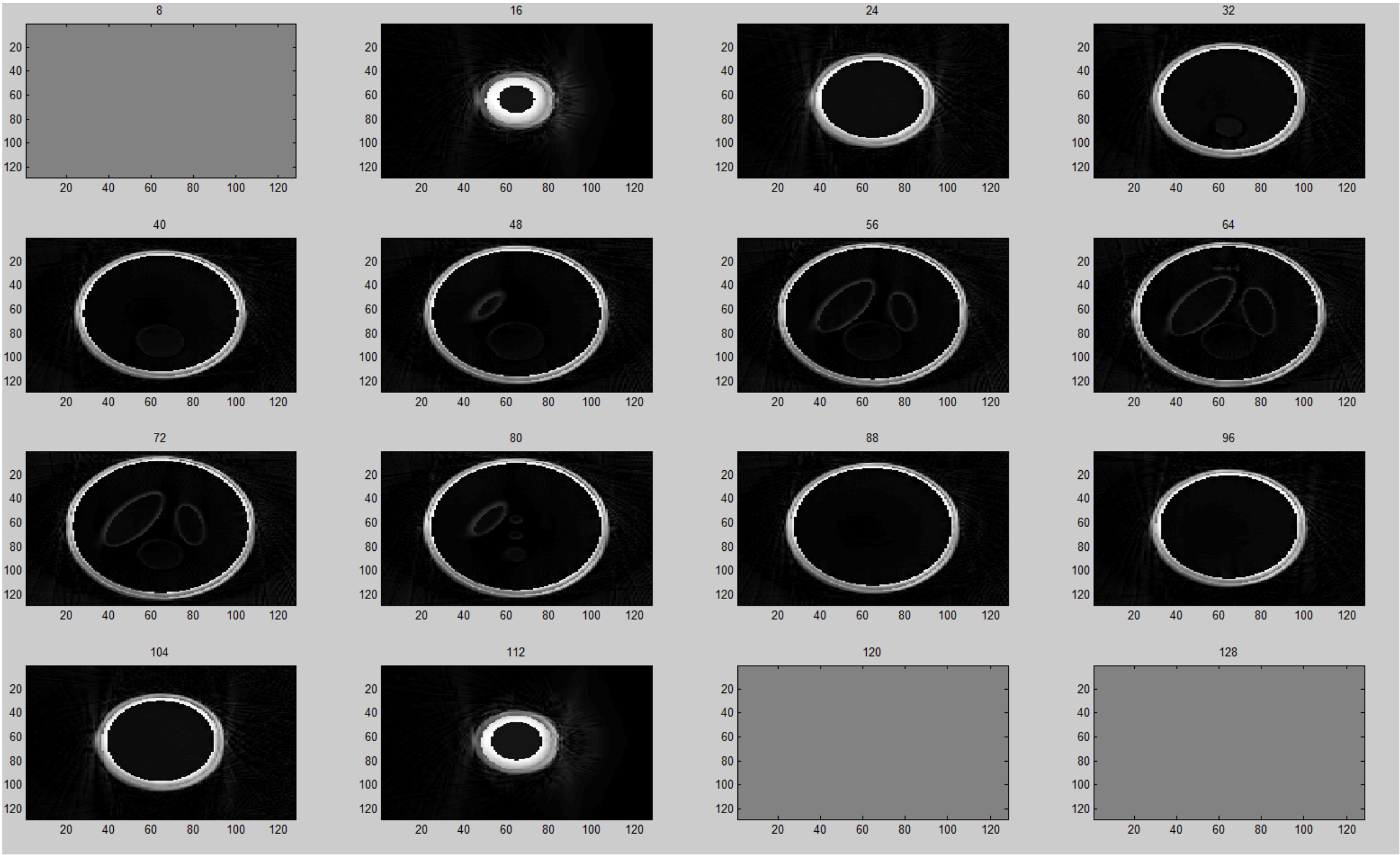

***Figure 3.21 The difference image between the reconstructed image by the FDK algorithm (from the projection data given by the NUFFT projector) and the original Shepp-Logan phantom***

From the results of the experiments, generally, we can see that the accuracy of the backprojector in reconstructing the contents of the phantom is acceptable intuitively, but it has problem in reconstructing the shell of the phantom.

The difference images shown in figure 3.15, 3.17 and 3.19 demonstrate that the NUFFT backprojector can give a reasonable approximation of inverting the NUFFT forward projector in spite of the possible discrepancy introduced by reversely using the resampling method for the forward direction. And on the other hand the NUFFT backprojector has potential to be treated as an alternative reconstruction algorithm for 3D cone beam direct reconstruction because the difference image results has large similarity.

From figure 3.19 and figure 3.21, we can see that the FDK algorithm is still more reliable in accuracy than the current stage NUFFT backprojector, especially at the shell of the phantom. On the other hand, note that these figures are the difference image between original 3D Shepp-Logan phantom and the 3D phantom image reconstructed from the projection data respectively given by the CT projector and the NUFFT projector. Because the FDK algorithm is a well-acknowledged cone beam reconstruction algorithm, and the FDK reconstruction results from the NUFFT projector' data and the CT projector's data is indistinguishable at least at the contents of the 3D Shepp-Logan phantom, we can conclude that the NUFFT forward projector designed in this project is able to get a very close approximation of the CT projector in terms of accuracy.

# 4. Summary and prospects

## *a) General discussion and summary of the algorithm designs in the project*

The main contain of this project is the innovative design of the NUFFT based forward projector which can provide computational efficiency improvement for iterative cone beam image reconstruction, while the NUFFT backprojector designed in section 2.3 is a by-product.

The difficulty and the key step of the NUFFT forward projector designed in this project is the 3D Sinogram resampling step. The section 2.1 gives a detail derivation of the NUFFT based resampling method, a complicated version (apply the first resampling method given by formula 2.11, involve a 3D IFFT and 3D NUFFT), and a simplified version (apply the second resampling method by formula 2.15, only use the interpolation matrix of the 3D NUFFT). The complexity analysis of the NUFFT projector and the CT projector is given by section 2.4.3, and the result of the analysis demonstrates that the NUFFT projector will have obvious computational advantage compared to the simplest version of CT projector based on the linear interpolation when the size of the CT imaging system described in that section is large. Typically when the 3D image is 512 cube-sized, the complicated version of NUFFT projector will be about 7 folds faster than the CT projector while the simplified version will be about 10 folds faster than the CT projector. Note that this complexity analysis is taken under the theoretical Sinogram sampling rates given by chapter 1.

The NUFFT based backprojector's design is in section 2.3. This is an attempt of designing a NUFFT based 3D filtered-backprojection algorithm for the circular cone beam geometry. But the 3D resampling method for the umbrella-to-radial direction has not been designed and implemented in this project yet, so temporary the resampling method designed for the forward projector is used reversely to take the place of the correct resampling method for the backward direction, and this will introduce some discrepancy in the reconstructed image. If the resampling method for the backward direction is successfully implemented in the future, the NUFFT backprojector will be a data-consistent 3D filtered backprojection algorithm.

## *b) Summary of the experiments and the further improvement of the implementations*

The experiments has been taken to test the rough performance of the forward projector and the backprojector in chapter 3. In spite of not fulfilling the theoretical sampling rates derived in chapter 1 for the 3D Sinogram but choosing a compromised version of the sampling rates at a cost of possible aliasing, the results of the experiments are really promising, especially for the joint experiment recorded in section 3.3, which involves the NUFFT forward projector (with the second resampling method) and backprojection algorithm, as well as the CT forward projector and the FDK backprojection algorithm. At first, the projection data sets given by both NUFFT projector and the CT projector are both treated as inputs. Next the NUFFT backprojection algorithm and the FDK algorithm are applied to reconstruct the 3D Shepp-Logan phantom from both sets of the projection data. This experiment given us a very convincing confirmation on the performance of the NUFFT forward projector since its reconstructed phantom by the FDK algorithm is very close to the original phantom and also the reconstructed phantom from the CT projector's data set. And on the other hand, the performance of the NUFFT backprojection is also demonstrated. From these good results given by the NUFFT based operators under a compromised 3D Sinogram sampling rates, we have enough reason to believe that in the further research when we implement them under the theoretical optimal sampling rate given by chapter 1, we will definitely get even better results.

From the experiments we can see that the first step of the future work should be adjusting the sampling number settings in the projectors to strictly follow the conclusion of the theoretical derivation in chapter 1 and apply these projectors again. The next step should be reviewing the details of the implementation and the Matlab code written for this project of the forward operator and find out the cause of the discrepancy in the projection data at the region outside the object and at the shell of the phantom. Another problem for the NUFFT projector implemented in this project is that the normalization of the output data is still needed (it means there is a scalar to be divided to scale the data to the normal level). Next the refined code for the resampling step is also needed to achieve the computational complexity promised in section 2.4 by formula (2.25) and (2.26). After fixing these problems, then we can apply it to the iterative cone beam algorithm and do experiments on it to see how it will perform.

For the backprojector's research, the next step should be correctly implement the resampling method in the backward direction. The backward resampling method given by C. Axelsson may be a reasonable choice of implementation, but it is a challenging task to code it in Matlab, so temporarily it has not yet been applied and tested. Next we can compare its accuracy with the FDK algorithm again. Because the NUFFT backprojector with the correct resampling method should be data-consistent, and the missing data problem for the circular cone beam CT will be reflected in the vacancy points occurred in the 3D (derivative of) Radon space, we can try to encode the prior knowledge information we have into these vacancies and see how it would improve the quality of the reconstructed image.

### *c) Some limitations and problems to be solved*

The NUFFT projector designed in this project certainly have some limitations to be investigated. From the results of the numerical experiment on determine the angular sampling number, we can see that for a 128 by 128 sized image (the images tested includes the slices in 3D Shepp-Logan, a real CT image and a random noise image), the number of the angular sampling we need to preserve the information of the image is lower than 256, but the figure shows that there exist a constant error which cannot be quenched by increasing the angular sampling, and in the case of the random noise image, the error become very obvious. This is because of the inherent limitation of the sampling pattern used in this project:

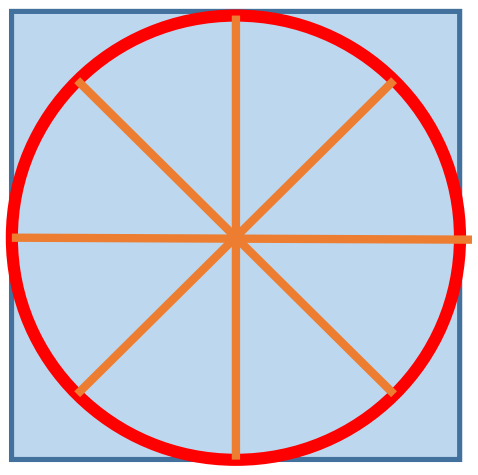

***Figure 4.1 The radial sampling pattern used in the project in the Fourier domain, we can see that the frequency components in the four corners are thrown away***

As shown in figure 4.1, the radial sampling pattern in this project will inherently ignore the frequency components at the four corners of the image's Fourier domain. With this we can explain the results of the experiment easily: the first three images are real-world images and the frequency component in the four corner is rare, but for the random noise image, this general property will not hold, so this constant error is relatively large.

Another limitation of this project is the sampling pattern we chose is the simplest one to analysis and implement, but it is an inefficient one. The figure 1.2 illustrate the uniform Sinogram sampling and its bowtie-shaped spectrum. We can see that the half of the space in the spectrum is not utilized.

In 2D case if we can use the sampling pattern illustrated by figure 1.2, the theoretical angular sampling number can be reduced to a half and the vacancy of the spectrum will be fill up and no aliasing will occur.

But the efficient way of sampling 3D Sinogram looks much more complicated to be found:

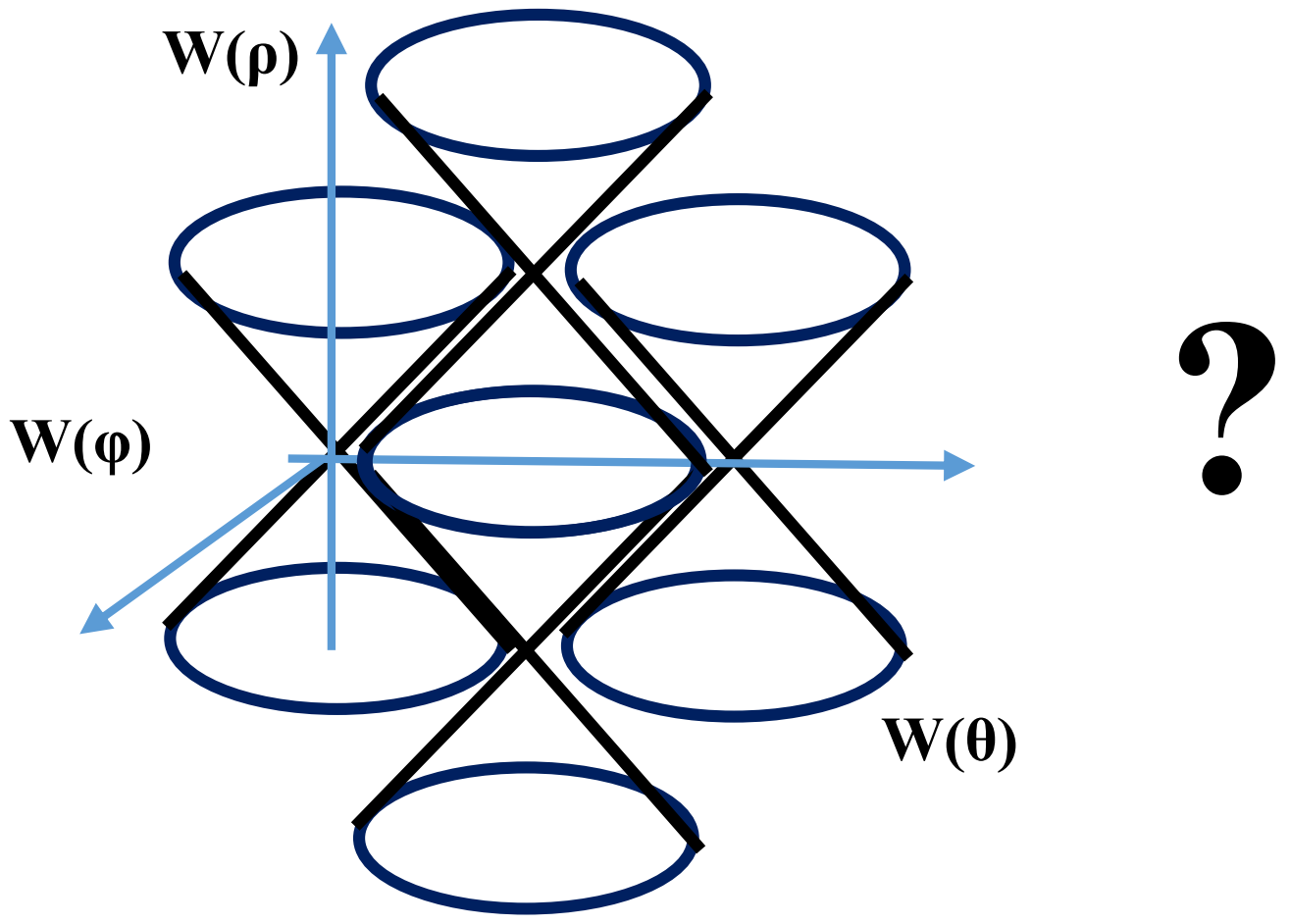

***Figure 4.2 The problem of sampling efficiently in the 3D Sinogram (Radon space)***

Finding an efficient way for sampling the 3D Radon space should also be the future direction of this research project.

There is also an important limitation of the resampling method derived in this project and strictly speaking it is not suitable for an efficient sampling pattern of 3D Sinogram (if there will be one in the future). As described in chapter 1 and the experiment 3.1, the sampling rate settings in this project is enough to preserve the information of the 3D image or its projection data sets, but unfortunately, the pre-assumption of the resampling method cannot hold because of the reduced angular sampling rate: it is originally designed for a 3D Sinogram which is uniformly sampled in terms of ($\rho$, $\varphi$, $\theta$) coordinate, and the sampling rate is adequate (Nyquist sampling) in each direction, a very inefficient sampling pattern. In the future step of the project, the resampling methods derived in this project should be investigated in detail and modified to be able to resample the efficient sampling pattern in Radon space without any possible aliasing.

***d) An alternative non-Fourier-based fast operator: possible benefit and limitation***

Besides the fast operators based on the Fourier Slice Theorem and NUFFT, there exists another stream of the non-Fourier based fast operator given by Y. Bressler [9] [10] [11], and this kind of the operator (FHP, fast hierarchical projector) is based on a totally different principle than the Fourier based operators but is able to get the same order of the complexity as them. This method can provide us an alternative and (possibly) more flexible way to perform the image-to-Radon transform, which is first step of the forward projector designed in this project, and sample the 3D image's Radon space more efficiently than using the Fourier Slice Theorem and have potential to reduce the complexity in this step.

But we also need to note that the FHP method may be inherently not suitable for cone beam geometry because the cone beam projection data is only related to the first order derivative of the Radon space, not the Radon space. This property will restrict any kinds of Radon-based data-consistent cone beam operator to only be operated in the derivative of Radon space. For Fourier based operator this is not a problem at all because doing derivative is very convenient through Fourier space, but it will definitely reduce the efficiency of the FHP algorithms. But anyway the FHP method is worth to be studied in the further steps of research of this project.

Generally speaking, the NUFFT based cone beam projector's research has a promising future because of its inherent advantage in computational efficiency, although the current stage designs, experiments and results still have some imperfection and need further study.

**Acknowledgements**

At the final moments of the MSc project my heart is filled with gratefulness towards my supervisor Prof. Mike Davies, who is full of passion for the research, without whose patience and guidance not only in the project but also in his course I could never have such solid knowledge about the project in this short period of time. And thanks for the great efforts, encouragements and supports from Perelli Alessandro and Jonathan Mason in the discussions and the advices about this Masters Report. It is such a great privilege and a precious experience to work with the team on this project.

*Junqi Tang*

*August 12, 2015 in the University of Edinburgh*